\documentclass[sigplan,screen]{acmart}

\usepackage{cleveref}
\usepackage{enumitem} 
\usepackage{caption} 
\usepackage{tikz}
\usepackage{pifont}
\usepackage{subcaption}

\makeatletter
\@ifundefined{lhs2tex.lhs2tex.sty.read}%
  {\@namedef{lhs2tex.lhs2tex.sty.read}{}%
   \newcommand\SkipToFmtEnd{}%
   \newcommand\EndFmtInput{}%
   \long\def\SkipToFmtEnd#1\EndFmtInput{}%
  }\SkipToFmtEnd

\newcommand\ReadOnlyOnce[1]{\@ifundefined{#1}{\@namedef{#1}{}}\SkipToFmtEnd}
\usepackage{amstext}
\usepackage{amssymb}
\usepackage{stmaryrd}
\DeclareFontFamily{OT1}{cmtex}{}
\DeclareFontShape{OT1}{cmtex}{m}{n}
  {<5><6><7><8>cmtex8
   <9>cmtex9
   <10><10.95><12><14.4><17.28><20.74><24.88>cmtex10}{}
\DeclareFontShape{OT1}{cmtex}{m}{it}
  {<-> ssub * cmtt/m/it}{}

\DeclareFontShape{OT1}{cmtt}{bx}{n}
  {<5><6><7><8>cmtt8
   <9>cmbtt9
   <10><10.95><12><14.4><17.28><20.74><24.88>cmbtt10}{}
\DeclareFontShape{OT1}{cmtex}{bx}{n}
  {<-> ssub * cmtt/bx/n}{}

\newcommand{\Conid}[1]{\mathit{#1}}
\newcommand{\Varid}[1]{\mathit{#1}}
\newcommand{\anonymous}{\kern0.06em \vbox{\hrule\@width.5em}}

\usepackage{polytable}

\@ifundefined{mathindent}%
  {\newdimen\mathindent\mathindent\leftmargini}%
  {}%

\def\resethooks{%
  \global\let\SaveRestoreHook\empty
  \global\let\ColumnHook\empty}
\newcommand*{\savecolumns}[1][default]%
  {\g@addto@macro\SaveRestoreHook{\savecolumns[#1]}}
\newcommand*{\restorecolumns}[1][default]%
  {\g@addto@macro\SaveRestoreHook{\restorecolumns[#1]}}
\newcommand*{\aligncolumn}[2]%
  {\g@addto@macro\ColumnHook{\column{#1}{#2}}}

\resethooks

\newcommand{\onelinecommentchars}{\quad-{}- }
\newcommand{\commentbeginchars}{\enskip\{-}
\newcommand{\commentendchars}{-\}\enskip}

\newcommand{\visiblecomments}{%
  \let\onelinecomment=\onelinecommentchars
  \let\commentbegin=\commentbeginchars
  \let\commentend=\commentendchars}

\newcommand{\invisiblecomments}{%
  \let\onelinecomment=\empty
  \let\commentbegin=\empty
  \let\commentend=\empty}

\visiblecomments

\newlength{\blanklineskip}
\newcommand{\hsindent}[1]{\quad}
\let\hspre\empty
\let\hspost\empty

\EndFmtInput
\makeatother
\ReadOnlyOnce{polycode.fmt}%
\makeatletter

\newcommand{\hsnewpar}[1]%
  {{\parskip=0pt\parindent=0pt\par\vskip #1\noindent}}

\newcommand{\hscodestyle}{}

\newcommand{\sethscode}[1]%
  {\expandafter\let\expandafter\hscode\csname #1\endcsname
   \expandafter\let\expandafter\endhscode\csname end#1\endcsname}

  {\par\noindent
   \advance\leftskip\mathindent
   \hscodestyle
   \let\\=\@normalcr
   \let\hspre\(\let\hspost\)%
   \pboxed}%
  {\endpboxed\)%
   \par\noindent
   \ignorespacesafterend}

  {\hsnewpar\abovedisplayskip
   \advance\leftskip\mathindent
   \hscodestyle
   \let\hspre\(\let\hspost\)%
   \pboxed}%
  {\endpboxed%
   \hsnewpar\belowdisplayskip
   \ignorespacesafterend}

  {\hsnewpar\abovedisplayskip
   \advance\leftskip\mathindent
   \hscodestyle
   \let\\=\@normalcr
   \(\pboxed}%
  {\endpboxed\)%
   \hsnewpar\belowdisplayskip
   \ignorespacesafterend}

\newcommand{\plainhs}{\sethscode{plainhscode}}

\plainhs

  {\hsnewpar\abovedisplayskip
   \advance\leftskip\mathindent
   \hscodestyle
   \let\\=\@normalcr
   \(\parray}%
  {\endparray\)%
   \hsnewpar\belowdisplayskip
   \ignorespacesafterend}

  {\parray}{\endparray}

  {\(\parray}{\endparray\)}

\def\codeframewidth{\arrayrulewidth}
\RequirePackage{calc}

  {\parskip=\abovedisplayskip\par\noindent
   \hscodestyle
   \arrayrulewidth=\codeframewidth
   \tabular{@{}|p{\linewidth-2\arraycolsep-2\arrayrulewidth-2pt}|@{}}%
   \hline\framedhslinecorrect\\{-1.5ex}%
   \let\endoflinesave=\\
   \let\\=\@normalcr
   \(\pboxed}%
  {\endpboxed\)%
   \framedhslinecorrect\endoflinesave{.5ex}\hline
   \endtabular
   \parskip=\belowdisplayskip\par\noindent
   \ignorespacesafterend}

\newcommand{\framedhslinecorrect}[2]%
  {#1[#2]}

  {\(\def\column##1##2{}%
   \let\>\undefined\let\<\undefined\let\\\undefined
   \newcommand\>[1][]{}\newcommand\<[1][]{}\newcommand\\[1][]{}%
   \def\fromto##1##2##3{##3}%
   }{\) }%

  {\let\orighscode=\hscode
   \let\origendhscode=\endhscode
   \def\endhscode{\def\hscode{\endgroup\def\@currenvir{hscode}\\}\begingroup}
   \orighscode\def\hscode{\endgroup\def\@currenvir{hscode}}}%
  {\origendhscode
   \global\let\hscode=\orighscode
   \global\let\endhscode=\origendhscode}%

\makeatother
\EndFmtInput
\long\def\ignore#1{}
\renewcommand\onelinecommentchars{-{}- }
\visiblecomments

\setcopyright{rightsretained}
\acmPrice{}
\acmDOI{10.1145/3609026.3609728}
\acmYear{2023}
\copyrightyear{2023}
\acmSubmissionID{icfpws23haskellmain-id51-p}
\acmISBN{979-8-4007-0298-3/23/09}
\acmConference[Haskell '23]{Proceedings of the 16th ACM SIGPLAN International Haskell Symposium}{September 8--9, 2023}{Seattle, WA, USA}
\acmBooktitle{Proceedings of the 16th ACM SIGPLAN International Haskell Symposium (Haskell '23), September 8--9, 2023, Seattle, WA, USA}
\received{2023-06-01}
\received[accepted]{2023-07-04}

\begin{document}

\title{An Exceptional Actor System (Functional Pearl)}

\author{Patrick Redmond}
\orcid{0000-0001-5702-0860}
\affiliation{
  \institution{University of California, Santa Cruz}
  \country{USA}
}
\author{Lindsey Kuper}
\orcid{0000-0002-1374-7715}
\affiliation{
  \institution{University of California, Santa Cruz}
  \country{USA}
}

\begin{abstract}
    The Glasgow Haskell Compiler is known for its feature-laden runtime system
    (RTS), which includes lightweight threads, asynchronous exceptions, and a
    slew of other features.
    Their combination is powerful enough that a programmer may
    complete the same task in many different ways --- some more advisable than
    others.

    We present a user-accessible actor framework hidden in plain sight within
    the RTS and demonstrate it on a classic example from the distributed
    systems literature.
    We then extend both the framework and example to the realm of dynamic
    types.
    Finally, we raise questions about how RTS features intersect and possibly
    subsume one another, and suggest that GHC can guide good practice by
    constraining the use of some features.
\end{abstract}



\maketitle

\section{Introduction}

Together with its runtime system (RTS), the Glasgow Haskell Compiler (GHC) is
the most commonly used implementation of Haskell \cite{fausak2022}.
The RTS is featureful and boasts support for lightweight threads, two kinds of
profiling, transactional memory, asynchronous exceptions, and more.
Combined with the \text{\ttfamily base} package, a programmer can get a lot
done without ever reaching into the extensive set of community packages on
Hackage.

In that spirit,
we noticed that there is nothing really stopping one from
abusing the tools \text{\ttfamily throwTo} and \text{\ttfamily catch}
to pass data between threads.
Any user-defined datatype can be made into an asynchronous exception.
Why not implement message-passing algorithms on that substrate?

We pursued this line of thought, and in this paper we present an actor
framework hidden just under the surface of the RTS.
The paper is organized as follows:
\begin{itemize}[leftmargin=1.5em]
    \item[--] \Cref{sec:background} provides a concise summary of asynchronous
    exceptions in GHC and the actor model of programming.

    \item[--] \Cref{sec:actor-framework} details the implementation of our
    actor framework. We first show how actors receive messages of a single
    type, and then extend the framework to support dynamically typed actors,
    which receive messages of more than one type.

    \item[--] \Cref{sec:ring-impl} shows an implementation of a classic
    protocol for leader election using our actor framework. We then extend the actors with an
    additional message type and behavior without changing the original
    implementation.

    \item[--] We reflect on whether this was a good idea in
    \Cref{sec:what-have-we-wrought},
    by considering the practicality and performance of our framework,
    and conclude in \Cref{sec:conclusion} that asynchronous exceptions
    might be more constrained.
\end{itemize}
This paper is a literate Haskell program.\footnote{
    We use \text{\ttfamily GHC~9\char46{}0\char46{}2} and \text{\ttfamily base\char45{}4\char46{}15\char46{}1\char46{}0}.
    Our actor framework imports \text{\ttfamily Control\char46{}Exception} and
    \text{\ttfamily Control\char46{}Concurrent}, and we use the extensions \text{\ttfamily NamedFieldPuns}
    and \text{\ttfamily DuplicateRecordFields} for convenience of presentation.
    The leader election example of \Cref{sec:ring-impl} additionally imports the module \text{\ttfamily System\char46{}Random}
    and uses the \text{\ttfamily ViewPatterns} extension.
    The appendices have other imports, which we do not describe here.
}

\ignore{
\begin{hscode}\SaveRestoreHook
\column{B}{@{}>{\hspre}l<{\hspost}@{}}%
\column{40}{@{}>{\hspre}l<{\hspost}@{}}%
\column{E}{@{}>{\hspre}l<{\hspost}@{}}%
\>[B]{}\mbox{\enskip\{-\# LANGUAGE NamedFieldPuns  \#-\}\enskip}{}\<[40]%
\>[40]{}\mbox{\onelinecomment  Section 2}{}\<[E]%
\\
\>[B]{}\mbox{\enskip\{-\# LANGUAGE DuplicateRecordFields  \#-\}\enskip}\mbox{\onelinecomment  Section 3.2}{}\<[E]%
\\
\>[B]{}\mbox{\enskip\{-\# LANGUAGE ViewPatterns  \#-\}\enskip}{}\<[40]%
\>[40]{}\mbox{\onelinecomment  Section 3.3}{}\<[E]%
\\[\blanklineskip]%
\>[B]{}\mbox{\onelinecomment  Section 2.1, 2.2}{}\<[E]%
\\
\>[B]{}\mathbf{import}\;\Conid{\Conid{Control}.Exception}\;(\Conid{Exception}\;(\mathinner{\ldotp\ldotp}),\Varid{throwTo},\Varid{catch},\Varid{mask\char95 }){}\<[E]%
\\
\>[B]{}\mathbf{import}\;\Conid{\Conid{Control}.Concurrent}\;(\Conid{ThreadId},\Varid{myThreadId},\Varid{threadDelay}){}\<[E]%
\\[\blanklineskip]%
\>[B]{}\mathbf{import}\;\Conid{\Conid{Control}.Exception}\;(\Varid{getMaskingState},\Conid{MaskingState}\;(\mathinner{\ldotp\ldotp})){}\<[E]%
\\[\blanklineskip]%
\>[B]{}\mbox{\onelinecomment  Section 2.3}{}\<[E]%
\\
\>[B]{}\mathbf{import}\;\Conid{\Conid{Control}.Exception}\;(\Conid{TypeError}\;(\mathinner{\ldotp\ldotp})){}\<[E]%
\\[\blanklineskip]%
\>[B]{}\mbox{\onelinecomment  Section 3.2}{}\<[E]%
\\
\>[B]{}\mathbf{import}\;\Conid{\Conid{Control}.Exception}\;(\Conid{SomeException}){}\<[E]%
\\
\>[B]{}\mathbf{import}\;\Conid{\Conid{Control}.Concurrent}\;(\Varid{forkIO},\Varid{killThread}){}\<[E]%
\\
\>[B]{}\mathbf{import}\;\Conid{\Conid{System}.Random}\;(\Conid{RandomGen},\Varid{randomR},\Varid{getStdRandom}){}\<[E]%
\\[\blanklineskip]%
\>[B]{}\mbox{\onelinecomment  Trace appendix}{}\<[E]%
\\
\>[B]{}\mathbf{import}\;\Conid{\Conid{System}.IO}\;(\Varid{hSetBuffering},\Varid{stdout},\Conid{BufferMode}\;(\mathinner{\ldotp\ldotp})){}\<[E]%
\\[\blanklineskip]%
\>[B]{}\mbox{\onelinecomment  Perf eval appendix}{}\<[E]%
\\
\>[B]{}\mathbf{import}\;\Conid{\Conid{Control}.Exception}\;(\Varid{assert}){}\<[E]%
\\
\>[B]{}\mathbf{import}\;\Conid{\Conid{System}.Environment}\;(\Varid{lookupEnv}){}\<[E]%
\\
\>[B]{}\mathbf{import}\;\Varid{qualified}\;\Conid{\Conid{Control}.\Conid{Concurrent}.Chan}\;\Varid{as}\;\Conid{Ch}{}\<[E]%
\\
\>[B]{}\mathbf{import}\;\Varid{qualified}\;\Conid{\Conid{Control}.\Conid{Concurrent}.MVar}\;\Varid{as}\;\Conid{Mv}{}\<[E]%
\\
\>[B]{}\mathbf{import}\;\Varid{qualified}\;\Conid{\Conid{Criterion}.Main}\;\Varid{as}\;\Conid{Cr}{}\<[E]%
\ColumnHook
\end{hscode}\resethooks
} 

\section{Brief background}
\label{sec:background}

In this section, we briefly review the status of asynchronous exceptions in GHC
(\Cref{subsec:async-exceptions}) and the actor model of programming
(\Cref{sec:actor-model}); readers already familiar with these topics may wish
to skip this section.
Readers unfamiliar with the behavior of \text{\ttfamily throwTo}, \text{\ttfamily catch}, or
\text{\ttfamily mask} from the \text{\ttfamily Control\char46{}Exception} module may wish to first scan the
documentation of \text{\ttfamily throwTo} in \citet{controlDotException}.

\subsection{Asynchronous exceptions in GHC}
\label{subsec:async-exceptions}

The Glasgow Haskell Compiler (GHC) is unusual in its support for
\emph{asynchronous exceptions}.
Unlike synchronous exceptions,
which are thrown as a result of executing code in the current thread,
asynchronous exceptions are thrown by threads distinct from the current one,
or by the RTS itself.
They are used to communicate conditions that may require the current thread to
terminate: thread cancellation, user interrupts, or memory limits.

Asynchronous exceptions allow syntactically-distant parts of a program
to interact in unexpected ways, much like mutable references.
A thread needs only the \text{\ttfamily ThreadId} of another
to throw a \text{\ttfamily ThreadKilled} exception to it.
The standard library function \text{\ttfamily killThread}
is even implemented as \text{\ttfamily \char40{}\char92{}x~\char45{}\char62{}~throwTo~x~ThreadKilled\char41{}}.\footnote{
    These identifiers are variously defined in \texttt{Control.Concurrent} and
    \texttt{Control.Exception} in \texttt{base-4.15.1.0}.
}
There is no permission or capability required to access this powerful feature.

Asynchronous exceptions are peculiar because they aren't constrained to their
stated purpose of ``signaling (or killing) one
thread by another'' \cite{marlow2001async}.
A thread may throw any exception to any thread for any reason.
This absence of restrictions means that standard exceptions may be reused for
any purpose, such as to extend greetings:
\text{\ttfamily \char40{}\char92{}x~\char45{}\char62{}~throwTo~x~\char36{}~AssertionFailed~\char34{}hello\char34{}\char41{}}.
Even
user-defined datatypes may be thrown as asynchronous exceptions by
declaring an empty instance of \text{\ttfamily Exception} \cite{marlow2006extensible}.
For example, with the declarations in \Cref{fig:greet}, it is possible to greet
in vernacular: \text{\ttfamily \char40{}\char92{}x~\char45{}\char62{}~throwTo~x~Hi\char41{}}.

Asynchronous exceptions may be caught by the receiving thread for
either cleanup or, surprisingly, recovery.
An example of recovery includes ``inform[ing] the program when memory is
running out [so] it can take remedial action'' \cite{marlow2001async}.
The ability to recover from a termination signal seems innocuous, but
it leaves asynchronous exceptions open to being repurposed.

\subsection{The actor model}
\label{sec:actor-model}

The actor model is a computational paradigm characterized by message passing.
\citet{hewitt1973actors} write that ``an actor can be thought of as a kind of
virtual processor that is never `busy' [in the sense that it cannot be sent a
message].''
In our setting, we interpret an actor to be a green thread\footnote{
    A \emph{green thread} (also ``lightweight thread'' or ``userspace thread'')
    is a thread not bound to an OS thread, but dynamically mapped to a CPU by a
    language-level scheduler.
    As opposed to heavier-weight OS threads, green threads simplify the implementation of a practical actor framework that supports large numbers of actors.
} with some state and an inbox.
When a message is received by an actor,
it is handled by that actor's \emph{intent function}.
An intent function may perform some actions:
send a message, update state, create a new actor, destroy an actor, or
terminate itself.
Unless terminated, the actor then waits to process the next message in its
inbox.
We will approximate this model with Haskell's asynchronous exceptions as the
mechanism for message passing.

More concretely, we think of an actor framework as
having the characteristics of a
\emph{concurrency-oriented programming language} (COPL),
a notion due to \citet{armstrong2003}.
After describing our framework, we will make the case (in \Cref{sec:almost-copl}) that it has many of the
characteristics of a COPL.
To summarize \citet{armstrong2003}, a COPL
(1) has processes,
(2) which are strongly isolated,
(3) with a unique hidden identifier,
(4) without shared state,
(5) that communicate via unreliable message passing,
and
(6) can detect when another process halts.
Additionally, 
(5a) message passing is asynchronous so that no stuck recipient may cause a sender to become stuck,
(5b) receiving a response is the only way to know that a prior message was sent,
and
(5c) messages between two processes obey FIFO ordering.
While an actor system within an instance of the RTS cannot satisfy all of these
requirements (e.g., termination of the main thread is not strongly isolated
from the child threads), we will show that our framework satisfies many requirements of
being a COPL with relatively little effort.

\begin{figure}
\begin{hscode}\SaveRestoreHook
\column{B}{@{}>{\hspre}l<{\hspost}@{}}%
\column{E}{@{}>{\hspre}l<{\hspost}@{}}%
\>[B]{}\mathbf{data}\;\Conid{Greet}\mathrel{=}\Conid{Hi}\mid \Conid{Hello}\;\mathbf{deriving}\;\Conid{Show}{}\<[E]%
\\
\>[B]{}\mathbf{instance}\;\Conid{Exception}\;\Conid{Greet}{}\<[E]%
\ColumnHook
\end{hscode}\resethooks
\caption{
    \text{\ttfamily Show} and \text{\ttfamily Exception} instances are all that is required to
    become an asynchronous exception.
}
\label{fig:greet}
\end{figure}

\section{Actor framework implementation}
\label{sec:actor-framework}

In our framework, an actor is a Haskell thread running a
provided main loop function.
The main loop function mediates message receipt and makes calls to a
user-defined intent function.
Here we describe the minimal abstractions around such threads that realize the
actor model.
These abstractions are so minimal as to seem unnecessary; we have sought to
keep them minimal to underscore our point.

\subsection{Sending (throwing) messages}
\label{sec:sending-throwing}

To send a message, we will throw an exception to the recipient's thread
identifier.
So that the recipient may respond, we define a self-addressed envelope data
type in \Cref{fig:envelope-and-intent} and declare the required instances.

\Cref{fig:static-impl} defines a send function, \text{\ttfamily sendStatic},
which reads the current thread identifier, constructs a self-addressed
envelope, and throws it to the specified recipient.
For the purpose of explication in this paper, it also prints an execution trace.

\subsection{Receiving (catching) messages}
\label{sec:receiving-catching}

An actor is defined by how it behaves in response to messages.
A user-defined intent function, with the type \text{\ttfamily Intent} shown in
\Cref{fig:envelope-and-intent},
encodes behavior as a state transition that takes a self-addressed envelope
argument.

Every actor thread will run a provided main loop function to manage message
receipt and processing.
The main loop function installs an exception handler to accumulate messages in
an inbox and calls a user-defined intent function on each.
\Cref{fig:static-impl} defines a main loop, \text{\ttfamily runStatic}, that
takes an \text{\ttfamily Intent} function and its initial state and does not return.
It masks asynchronous exceptions so they will only be raised at well-defined
points within the loop: during \text{\ttfamily threadDelay} or possibly during the
\text{\ttfamily Intent} function.

The loop in \Cref{fig:static-impl} has two pieces of state: that of the intent
function, and an inbox of messages to be processed.
The loop body is divided roughly into three cases by an exception
handler and a case-split on the inbox list:
\begin{enumerate}[leftmargin=2em]
    \item If the inbox is empty, sleep for an arbitrary length of time and then
    recurse on the unchanged actor state and the empty inbox.
    
    \item If the inbox has a message, call the intent function and recurse on
    the updated actor state and the remainder of the inbox.

    \item If, during cases (1) or (2), an \text{\ttfamily Envelope} exception is received,
    recurse on the unchanged actor state and an inbox with the new envelope
    appended to the end.
\end{enumerate}
In the normal course of things, an actor will start with an empty inbox and go
to sleep.
If a message is received during sleep, the actor will wake (because
\text{\ttfamily threadDelay} is defined to be \emph{interruptible}), add the message to
its inbox, and recurse.
On the next loop iteration, the actor will process that message and once again
have an empty inbox.
Exceptions are masked (using \texttt{mask\_}\footnote{
    It is good practice to use \texttt{mask} instead of \texttt{mask\_}, and
    ``restore'' the prior masking state of the context before calling a
    user-defined callback function.
    Such functions may be written with the expectation to catch asynchronous
    exceptions, for reasons mentioned in \Cref{subsec:async-exceptions} or
    \citet{marlow2001async}.
    For our purpose here, \texttt{mask\_} is acceptable.
}) outside of interruptible actions so that the bookkeeping
of recursing with updated state through the loop is not disrupted.

\paragraph{Unsafety}

Before moving forward, let us acknowledge that this is \emph{not safe}.
An exception may arrive while executing the intent function.
Despite our use of \texttt{mask\_},
if the intent function executes an interruptible action, then
it will be preempted.
In this case the intent function's work will be unfinished.
Without removing the message currently being processed, the loop
will continue on an inbox extended with the new message.
The next iteration will begin by processing the same message that the preempted
iteration was, effecting a double-send.

To avoid the possibility of a double-send, a careful implementor of an actor
program might follow the documented recommendations for code in the presence of
asynchronous exceptions:
use software transactional memory (STM),
avoid interruptible actions,
or apply \text{\ttfamily uninterruptibleMask}.
However, recall that message sends are implemented with \text{\ttfamily throwTo}, which is
``\emph{always} interruptible, even if it does not actually block''
\cite{controlDotException}.
A solution is obtained ``by forking a new thread'' \cite{marlow2001async} each
time we run an intent function, but this sacrifices serializable
executions --- an actor must be safe to run concurrently with itself.
We opt for the simple presentation in \Cref{fig:static-impl}
and recommend users write idempotent intent functions.

\begin{figure}
\raggedright
\begin{hscode}\SaveRestoreHook
\column{B}{@{}>{\hspre}l<{\hspost}@{}}%
\column{5}{@{}>{\hspre}l<{\hspost}@{}}%
\column{E}{@{}>{\hspre}l<{\hspost}@{}}%
\>[B]{}\mathbf{data}\;\Conid{Envelope}\;\Varid{a}\mathrel{=}\Conid{Envelope}\;\!\{\Varid{sender}\mathbin{::}\Conid{ThreadId},\Varid{message}\mathbin{::}\Varid{a}\mskip1.5mu\}{}\<[E]%
\\
\>[B]{}\hsindent{5}{}\<[5]%
\>[5]{}\mathbf{deriving}\;\Conid{Show}{}\<[E]%
\\[\blanklineskip]%
\>[B]{}\mathbf{instance}\;\Conid{Exception}\;\Varid{a}\Rightarrow \Conid{Exception}\;(\Conid{Envelope}\;\Varid{a}){}\<[E]%
\\[\blanklineskip]%
\>[B]{}\mathbf{type}\;\Conid{Intent}\;\Varid{st}\;\Varid{msg}\mathrel{=}\Varid{st}\to \Conid{Envelope}\;\Varid{msg}\to \Conid{IO}\;\Varid{st}{}\<[E]%
\ColumnHook
\end{hscode}\resethooks
\caption{
    Message values are contained in a self-addressed envelope.
    Actor behavior is encoded as a transition system.
}
\label{fig:envelope-and-intent}
\end{figure}

\begin{figure}
\raggedright
\begin{hscode}\SaveRestoreHook
\column{B}{@{}>{\hspre}l<{\hspost}@{}}%
\column{3}{@{}>{\hspre}l<{\hspost}@{}}%
\column{5}{@{}>{\hspre}l<{\hspost}@{}}%
\column{9}{@{}>{\hspre}l<{\hspost}@{}}%
\column{13}{@{}>{\hspre}l<{\hspost}@{}}%
\column{17}{@{}>{\hspre}l<{\hspost}@{}}%
\column{21}{@{}>{\hspre}l<{\hspost}@{}}%
\column{27}{@{}>{\hspre}l<{\hspost}@{}}%
\column{E}{@{}>{\hspre}l<{\hspost}@{}}%
\>[B]{}\Varid{sendStatic}\mathbin{::}\Conid{Exception}\;\Varid{a}\Rightarrow \Conid{ThreadId}\to \Varid{a}\to \Conid{IO}\;(){}\<[E]%
\\
\>[B]{}\Varid{sendStatic}\;\Varid{recipient}\;\Varid{message}\mathrel{=}\mathbf{do}{}\<[E]%
\\
\>[B]{}\hsindent{5}{}\<[5]%
\>[5]{}\Varid{sender}\leftarrow \Varid{myThreadId}{}\<[E]%
\\
\>[B]{}\hsindent{5}{}\<[5]%
\>[5]{}\Varid{putStrLn}\;(\Varid{show}\;\Varid{sender}\mathbin{+\hspace{-0.2em}+}\text{\ttfamily \char34 ~send~\char34}\mathbin{+\hspace{-0.2em}+}\Varid{show}\;\Varid{message}{}\<[E]%
\\
\>[5]{}\hsindent{22}{}\<[27]%
\>[27]{}\mathbin{+\hspace{-0.2em}+}\text{\ttfamily \char34 ~to~\char34}\mathbin{+\hspace{-0.2em}+}\Varid{show}\;\Varid{recipient}){}\<[E]%
\\
\>[B]{}\hsindent{5}{}\<[5]%
\>[5]{}\Varid{throwTo}\;\Varid{recipient}\;\Conid{Envelope}\;\!\{\Varid{sender},\Varid{message}\mskip1.5mu\}{}\<[E]%
\\[\blanklineskip]%
\>[B]{}\Varid{runStatic}\mathbin{::}\Conid{Exception}\;\Varid{a}\Rightarrow \Conid{Intent}\;\Varid{s}\;\Varid{a}\to \Varid{s}\to \Conid{IO}\;(){}\<[E]%
\\
\>[B]{}\Varid{runStatic}\;\Varid{intent}\;\Varid{initialState}\mathrel{=}\Varid{mask\char95 }\mathbin{\$}\Varid{loop}\;(\Varid{initialState},[\mskip1.5mu]){}\<[E]%
\\
\>[B]{}\hsindent{3}{}\<[3]%
\>[3]{}\mathbf{where}{}\<[E]%
\\
\>[3]{}\hsindent{2}{}\<[5]%
\>[5]{}\Varid{loop}\;(\Varid{state},\Varid{inbox})\mathrel{=}{}\<[E]%
\\
\>[5]{}\hsindent{4}{}\<[9]%
\>[9]{}\Varid{catch}\;{}\<[E]%
\\
\>[9]{}\hsindent{4}{}\<[13]%
\>[13]{}(\mathbf{case}\;\Varid{inbox}\;\mathbf{of}{}\<[E]%
\\
\>[13]{}\hsindent{4}{}\<[17]%
\>[17]{}[\mskip1.5mu]\to \Varid{threadDelay}\;\mathrm{60000000}\hfill(1){}\<[E]%
\\
\>[17]{}\hsindent{4}{}\<[21]%
\>[21]{}\mathbin{>\hspace{-0.4em}>}\Varid{return}\;(\Varid{state},\Varid{inbox}){}\<[E]%
\\
\>[13]{}\hsindent{4}{}\<[17]%
\>[17]{}\Varid{x}\mathbin{:}\Varid{xs}\to {}\<[E]%
\\
\>[17]{}\hsindent{4}{}\<[21]%
\>[21]{}(,\!)\mathbin{\langle\$\rangle}\Varid{intent}\;\Varid{state}\;\Varid{x}\mathbin{\langle*\rangle}\Varid{return}\;\Varid{xs})\;\hfill(2){}\<[E]%
\\
\>[9]{}\hsindent{4}{}\<[13]%
\>[13]{}(\lambda \Varid{e}\mathord{@}\Conid{Envelope}\;\!\{\mskip1.5mu\}\to {}\<[E]%
\\
\>[13]{}\hsindent{4}{}\<[17]%
\>[17]{}\Varid{return}\;(\Varid{state},\Varid{inbox}\mathbin{+\hspace{-0.2em}+}[\Varid{e}\mskip1.5mu]))\hfill(3){}\<[E]%
\\
\>[5]{}\hsindent{4}{}\<[9]%
\>[9]{}\mathbin{>\hspace{-0.4em}>\hspace{-0.3em}=}\Varid{loop}{}\<[E]%
\ColumnHook
\end{hscode}\resethooks
\caption{
    Message sends are implemented by throwing an exception.
    Actor threads run a main loop to receive messages.
}
\label{fig:static-impl}
\end{figure}

\subsection{Dynamic types}
\label{sec:dynamic-types}

The actor main loop in \Cref{fig:static-impl} constrains an actor
thread to handle messages of a single type.
An envelope containing the wrong message type will not be caught by the
exception handler, causing the receiving actor to crash.
We think the recipient should not crash when another actor sends an incorrect
message.\footnote{
    Sending a message not handled by the recipient is like calling a function
    with wrong argument types, which would cause the thread to crash in a
    dynamically typed language. However, here both caller and callee are
    persistent, and we choose to locate the mistake in the caller.
}

In this section, we correct this issue by extending the framework to support
actors that may receive messages of different types.
With this extension, our framework could be thought of as dynamically typed in
the sense that a single actor can process multiple message types.
This is similar to the dynamic types support in the
\text{\ttfamily Data\char46{}Dynamic} module.

Furthermore, any actor may be extended by wrapping it (``has-a'' style) with an
actor that uses a distinct message type and branches on the type of a received
message, delegating to the wrapped actor where desired.\footnote{
    It is not sufficient to wrap a message type in a sum and write an actor
    that takes the sum as its message.
    Such an actor will fail to receive messages sent as the un-wrapped type.
    To correct for this, one would need to change existing actors to wrap their
    outgoing messages in the sum.
    \Cref{sec:dynamic-types} generalizes this
    correction without requiring changes to existing actors.
}
It may seem natural to encapsulate such actor-wrapping in combinators that
generalize the patterns by which an actor is given additional behavior.
However, here our goal is not to lean into the utility of a dynamically
typed actor framework, but to point out how little scaffolding is required to
obtain one from the RTS.

\subsubsection{Sending dynamic messages}

Instead of sending an \text{\ttfamily Envelope}
of some application-specific message type
we convert messages to the ``any type''
in Haskell's exception hierarchy,
\text{\ttfamily SomeException} \cite{marlow2006extensible}.
\Cref{fig:dyn-impl} defines a new \text{\ttfamily send} function that converts messages,
so that all inflight messages will have the type \text{\ttfamily Envelope~SomeException}.

\subsubsection{Receiving dynamic messages}
\label{sec:dynamic-recv-loop}

On the receiving side, messages must now be downcast to the \text{\ttfamily Intent}
function's message type.
This is an opportunity to treat messages of the wrong type specially.
In \Cref{fig:dyn-impl} we define a new main loop, \text{\ttfamily runDyn},
that lifts any intent function to one that can
receive envelopes containing \text{\ttfamily SomeException}.
If the message downcast fails, instead of the recipient crashing, it performs a
``return to sender.''
Specifically, it throws an exception (not an envelope) using the built-in
\text{\ttfamily TypeError} exception.\footnote{
    The extensions \texttt{ScopedTypeVariables}, \texttt{TypeApplications}, and
    the function \texttt{Data.Typeable.typeOf} can be used to construct a
    helpful type error message for debugging actor programs.
}

These changes do not directly empower actor intent functions to
deal with messages of different types.
We have only removed
application-specific type parameters from envelopes.
Actors intending to receive messages of different types will do so by
downcasting from \text{\ttfamily SomeException} themselves.
Such actors will use an intent function handling messages of type
\text{\ttfamily SomeException}.
%
We will see an example of this usage pattern in \Cref{sec:dyn-ring}.

\begin{figure}
\raggedright
\begin{hscode}\SaveRestoreHook
\column{B}{@{}>{\hspre}l<{\hspost}@{}}%
\column{3}{@{}>{\hspre}l<{\hspost}@{}}%
\column{5}{@{}>{\hspre}l<{\hspost}@{}}%
\column{9}{@{}>{\hspre}l<{\hspost}@{}}%
\column{13}{@{}>{\hspre}l<{\hspost}@{}}%
\column{17}{@{}>{\hspre}l<{\hspost}@{}}%
\column{E}{@{}>{\hspre}l<{\hspost}@{}}%
\>[B]{}\Varid{send}\mathbin{::}\Conid{Exception}\;\Varid{a}\Rightarrow \Conid{ThreadId}\to \Varid{a}\to \Conid{IO}\;(){}\<[E]%
\\
\>[B]{}\Varid{send}\;\Varid{recipient}\mathrel{=}\Varid{sendStatic}\;\Varid{recipient}\mathbin{\circ}\Varid{toException}{}\<[E]%
\\[\blanklineskip]%
\>[B]{}\Varid{runDyn}\mathbin{::}\Conid{Exception}\;\Varid{a}\Rightarrow \Conid{Intent}\;\Varid{s}\;\Varid{a}\to \Varid{s}\to \Conid{IO}\;(){}\<[E]%
\\
\>[B]{}\Varid{runDyn}\;\Varid{intentStatic}\mathrel{=}\Varid{runStatic}\;\Varid{intentDyn}{}\<[E]%
\\
\>[B]{}\hsindent{3}{}\<[3]%
\>[3]{}\mathbf{where}{}\<[E]%
\\
\>[3]{}\hsindent{2}{}\<[5]%
\>[5]{}\Varid{intentDyn}\;\Varid{state}\;\Varid{e}\mathord{@}\Conid{Envelope}\;\!\{\Varid{sender},\Varid{message}\mskip1.5mu\}\mathrel{=}{}\<[E]%
\\
\>[5]{}\hsindent{4}{}\<[9]%
\>[9]{}\mathbf{case}\;\Varid{fromException}\;\Varid{message}\;\mathbf{of}{}\<[E]%
\\
\>[9]{}\hsindent{4}{}\<[13]%
\>[13]{}\Conid{Just}\;\Varid{m}\to \Varid{intentStatic}\;\Varid{state}\;\Varid{e}\;\!\{\Varid{message}\mathrel{=}\Varid{m}\mskip1.5mu\}{}\<[E]%
\\
\>[9]{}\hsindent{4}{}\<[13]%
\>[13]{}\Conid{Nothing}{}\<[E]%
\\
\>[13]{}\hsindent{4}{}\<[17]%
\>[17]{}\to \Varid{throwTo}\;\Varid{sender}\;(\Conid{TypeError}\;\text{\ttfamily \char34 ...\char34}){}\<[E]%
\\
\>[13]{}\hsindent{4}{}\<[17]%
\>[17]{}\mathbin{>\hspace{-0.4em}>}\Varid{return}\;\Varid{state}{}\<[E]%
\ColumnHook
\end{hscode}\resethooks
\caption{
    The dynamically typed framework
    upcasts before sending
    and downcasts before processing.
}
\label{fig:dyn-impl}
\end{figure}

\subsection{Safe initialization}
\label{sec:safe-fork}

When creating an actor thread, it is important that no exception arrive before
the actor main loop (\text{\ttfamily runStatic} in \Cref{fig:static-impl})
installs its exception handler.
If this happened, the exception would cause the newly created thread to die.
To avoid this, the fork prior to entering the main loop must be
masked (in addition to the mask within the main loop).

\Cref{fig:run} defines the main loop wrapper we will use for examples in
\Cref{sec:ring-impl}.
It performs a best-effort check and issues a helpful reminder to mask the
creation of actor threads.\footnote{
    We do not define a wrapper around \texttt{forkIO} to perform this masking
    because actors that perform initialization steps can currently do so
    before calling \texttt{run}. \Cref{sec:main2-init} is an example of this.
}

\begin{figure}
\raggedright
\begin{hscode}\SaveRestoreHook
\column{B}{@{}>{\hspre}l<{\hspost}@{}}%
\column{5}{@{}>{\hspre}l<{\hspost}@{}}%
\column{9}{@{}>{\hspre}l<{\hspost}@{}}%
\column{E}{@{}>{\hspre}l<{\hspost}@{}}%
\>[B]{}\Varid{run}\mathbin{::}\Conid{Exception}\;\Varid{a}\Rightarrow \Conid{Intent}\;\Varid{s}\;\Varid{a}\to \Varid{s}\to \Conid{IO}\;(){}\<[E]%
\\
\>[B]{}\Varid{run}\;\Varid{intent}\;\Varid{state}\mathrel{=}\mathbf{do}{}\<[E]%
\\
\>[B]{}\hsindent{5}{}\<[5]%
\>[5]{}\Varid{ms}\leftarrow \Varid{getMaskingState}{}\<[E]%
\\
\>[B]{}\hsindent{5}{}\<[5]%
\>[5]{}\mathbf{case}\;\Varid{ms}\;\mathbf{of}{}\<[E]%
\\
\>[5]{}\hsindent{4}{}\<[9]%
\>[9]{}\Conid{MaskedInterruptible}\to \Varid{runDyn}\;\Varid{intent}\;\Varid{state}{}\<[E]%
\\
\>[5]{}\hsindent{4}{}\<[9]%
\>[9]{}\anonymous \to \Varid{error}\;\text{\ttfamily \char34 mask~the~forking~of~actor~threads\char34}{}\<[E]%
\ColumnHook
\end{hscode}\resethooks
\caption{Remind users to prevent initialization errors by masking forks.}
\label{fig:run}
\end{figure}

\section{Example: Ring leader election}
\label{sec:ring-impl}

The problem of \emph{ring leader election} is to designate one node
among a network of communicating nodes organized in a ring topology.
Each node has a unique identity, and identities are totally ordered.
Nodes know their immediate successor, or ``next'' node, but do not know the
number or identities of the other nodes in the ring.
A correct solution will result in exactly one node being designated the leader.
This classic problem from the distributed
systems literature serves to illustrate our actor framework,
despite leader election being unnecessary in the context of threads in a process.

\begin{figure}
\begin{tikzpicture}[scale=1.5]

\def\radius{2cm}
\def\bounce{18pt}
\def\pad{10pt}
\def\ct{7}
\def\noderad{8pt}
\coordinate (center) at (1,1);

\draw (center) circle[radius=\radius];

\fill[white] (center) +(0/\ct*360:\radius) circle[radius=\noderad];
\fill[white] (center) +(1/\ct*360:\radius) circle[radius=\noderad];
\fill[white] (center) +(2/\ct*360:\radius) circle[radius=\noderad];
\fill[white] (center) +(3/\ct*360:\radius) circle[radius=\noderad];
\fill[white] (center) +(4/\ct*360:\radius) circle[radius=\noderad];
\fill[white] (center) +(5/\ct*360:\radius) circle[radius=\noderad];
\fill[white] (center) +(6/\ct*360:\radius) circle[radius=\noderad];

\draw (center) +(0/\ct*360:\radius) circle[radius=\noderad] node {6};
\draw (center) +(1/\ct*360:\radius) circle[radius=\noderad] node {2};
\draw (center) +(2/\ct*360:\radius) circle[radius=\noderad] node {7};
\draw (center) +(3/\ct*360:\radius) circle[radius=\noderad] node {5};
\draw (center) +(4/\ct*360:\radius) circle[radius=\noderad] node {3};
\draw (center) +(5/\ct*360:\radius) circle[radius=\noderad] node {1};
\draw (center) +(6/\ct*360:\radius) circle[radius=\noderad] node {4};


\draw[thick,->]
          (center) +(3.1/\ct*360:\radius+\pad)
       .. controls +(4/\ct*360:\bounce)
               and +(3/\ct*360:\bounce)
                .. +(3.9/\ct*360:\radius+\pad);

\draw[thick,->]
          (center) +(4.1/\ct*360:\radius+\pad)
       .. controls +(5/\ct*360:\bounce)
               and +(4/\ct*360:\bounce)
                .. +(4.9/\ct*360:\radius+\pad);

\draw[thick,->]
          (center) +(5.1/\ct*360:\radius+\pad)
       .. controls +(6/\ct*360:\bounce)
               and +(5/\ct*360:\bounce)
                .. +(5.9/\ct*360:\radius+\pad);


\draw (center) +(3.5/\ct*360:\radius+\bounce+2mm) node {5};
\draw (center) +(4.5/\ct*360:\radius+\bounce+2mm) node {5};
\draw (center) +(5.5/\ct*360:\radius+\bounce+2mm) node {5};


\draw (center) +(4/\ct*360:\radius+2*\noderad) node {\ding{51}}; 
\draw (center) +(5/\ct*360:\radius+2*\noderad) node {\ding{51}}; 
\draw (center) +(6/\ct*360:\radius+2*\noderad) node {\ding{51}}; 



\draw[thick,->]
          (center) +(0.1/\ct*360:\radius+\pad)
       .. controls +(1/\ct*360:\bounce)
               and +(0/\ct*360:\bounce)
                .. +(0.9/\ct*360:\radius+\pad);

\draw[thick,->]
          (center) +(1.1/\ct*360:\radius+\pad)
       .. controls +(2/\ct*360:\bounce)
               and +(1/\ct*360:\bounce)
                .. +(1.9/\ct*360:\radius+\pad);


\draw (center) +(0.5/\ct*360:\radius+\bounce+2mm) node {6};
\draw (center) +(1.5/\ct*360:\radius+\bounce+2mm) node {6};


\draw (center) +(1/\ct*360:\radius+2*\noderad) node {\ding{51}}; 
\draw (center) +(2/\ct*360:\radius+2*\noderad) node {\ding{55}}; 

\end{tikzpicture}
\caption{
    In-progress ring leader election with seven nodes
    (\citeauthor{chang1979decentralextrema}'
    \citeyear{chang1979decentralextrema} solution).
    The node identities are unique and randomly distributed.
    Two nomination chains are shown:
    Node 5 nominated itself and was accepted by nodes 3, 1, and 4; next node 4
    will nominate 5 to node 6 (who will reject it).
    Concurrently, node 6 nominated itself and was accepted by node 2 but
    rejected by node 7.
    For this election to result in a leader, node 7 must nominate itself.
}
\label{fig:ring-election-visual}
\end{figure}
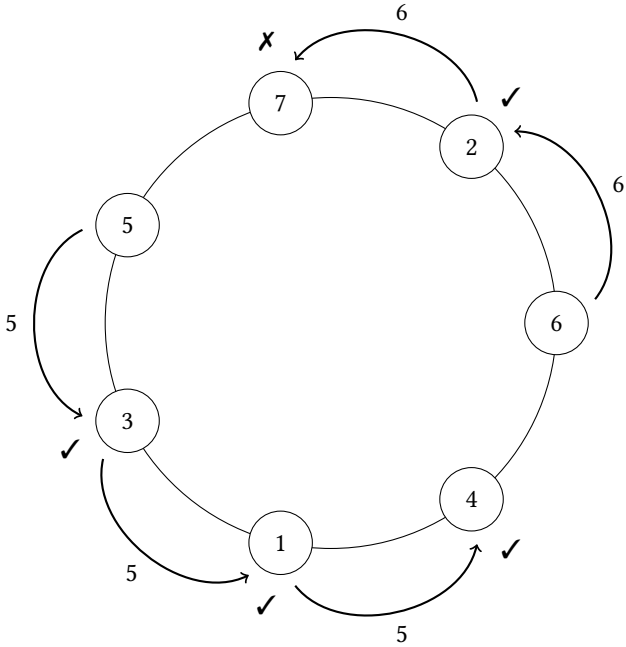

\citet{chang1979decentralextrema} describe a solution to the ring leader election problem that begins with every
node sending a message to its successor to nominate itself as the leader
(\Cref{fig:ring-election-visual}).
Upon receiving a nomination,
a node forwards the nomination to its successor
if the identity of the nominee is
greater than its own identity.
Otherwise, the nomination is ignored.
We implement and extend that solution below.

\subsection{Implementing a leader election}

Each node begins uninitialized, and later becomes a member of the ring when
it learns the identity of its successor.
To represent this we define two constructors in \Cref{fig:node-types} for node
state type, \text{\ttfamily Node}.

\noindent
Three messages (also defined in \Cref{fig:node-types} as type, \text{\ttfamily Msg}) will
be used to run the election:
\begin{itemize}[leftmargin=1.5em]
    \item[--] \text{\ttfamily Init}: After creating nodes, the main thread initializes
    the ring by informing each node of its successor.
    \item[--] \text{\ttfamily Start}: The main thread rapidly instructs every node to start
    the leader election.
    \item[--] \text{\ttfamily Nominate}: The nodes carry out the election by sending and
    receiving nominations.
\end{itemize}
\begin{figure}
\raggedright
\begin{hscode}\SaveRestoreHook
\column{B}{@{}>{\hspre}l<{\hspost}@{}}%
\column{5}{@{}>{\hspre}l<{\hspost}@{}}%
\column{E}{@{}>{\hspre}l<{\hspost}@{}}%
\>[B]{}\mathbf{data}\;\Conid{Node}\mathrel{=}\Conid{Uninitialized}\mid \Conid{Member}\;\!\{\Varid{next}\mathbin{::}\Conid{ThreadId}\mskip1.5mu\}{}\<[E]%
\\[\blanklineskip]%
\>[B]{}\mathbf{data}\;\Conid{Msg}{}\<[E]%
\\
\>[B]{}\hsindent{5}{}\<[5]%
\>[5]{}\mathrel{=}\Conid{Init}\;\!\{\Varid{next}\mathbin{::}\Conid{ThreadId}\mskip1.5mu\}{}\<[E]%
\\
\>[B]{}\hsindent{5}{}\<[5]%
\>[5]{}\mid \Conid{Start}{}\<[E]%
\\
\>[B]{}\hsindent{5}{}\<[5]%
\>[5]{}\mid \Conid{Nominate}\;\!\{\Varid{nominee}\mathbin{::}\Conid{ThreadId}\mskip1.5mu\}{}\<[E]%
\\
\>[B]{}\hsindent{5}{}\<[5]%
\>[5]{}\mathbf{deriving}\;\Conid{Show}{}\<[E]%
\\[\blanklineskip]%
\>[B]{}\mathbf{instance}\;\Conid{Exception}\;\Conid{Msg}{}\<[E]%
\ColumnHook
\end{hscode}\resethooks
\caption{
    Election nodes can be in one of two states,
    and they accept three different messages.
}
\label{fig:node-types}
\end{figure}

\subsubsection{Election termination}
\label{sec:election-termination}
The node with the greatest identity that nominates itself will eventually
receive its own nomination after it has circulated the entire ring.
That same node will ignore every other nomination.
Therefore the algorithm will terminate because node identities are unique
and only one nomination can circumnavigate the ring.\footnote{
    In the context of this paper, termination is guaranteed because we have
    reliable message passing (see \Cref{sec:almost-copl}).
    In the context of a distributed system, with unreliable message passing, it
    is possible that no nomination makes it all the way around the ring.
    In such a situation, the algorithm could terminate without a
    winner.
}

\subsubsection{Node-actor behavior}
\label{sec:ring-intent-fun}

The intent function for a node actor will have state of type \text{\ttfamily Node} and
receive messages of type \text{\ttfamily Msg}, as defined in \Cref{fig:node-types}.
We show its implementation and describe each case below.
\begin{hscode}\SaveRestoreHook
\column{B}{@{}>{\hspre}l<{\hspost}@{}}%
\column{E}{@{}>{\hspre}l<{\hspost}@{}}%
\>[B]{}\Varid{node}\mathbin{::}\Conid{Intent}\;\Conid{Node}\;\Conid{Msg}{}\<[E]%
\ColumnHook
\end{hscode}\resethooks
When an uninitialized node receives an \text{\ttfamily Init} message, it becomes a member
of the ring and remembers its successor.
\begin{hscode}\SaveRestoreHook
\column{B}{@{}>{\hspre}l<{\hspost}@{}}%
\column{3}{@{}>{\hspre}l<{\hspost}@{}}%
\column{5}{@{}>{\hspre}l<{\hspost}@{}}%
\column{E}{@{}>{\hspre}l<{\hspost}@{}}%
\>[B]{}\Varid{node}\;\Conid{Uninitialized}\;{}\<[E]%
\\
\>[B]{}\hsindent{3}{}\<[3]%
\>[3]{}\Conid{Envelope}\;\!\{\Varid{message}\mathrel{=}\Conid{Init}\;\!\{\Varid{next}\mskip1.5mu\}\mskip1.5mu\}\mathrel{=}\mathbf{do}{}\<[E]%
\\
\>[3]{}\hsindent{2}{}\<[5]%
\>[5]{}\Varid{return}\;\Conid{Member}\;\!\{\Varid{next}\mskip1.5mu\}{}\<[E]%
\ColumnHook
\end{hscode}\resethooks
When a member of the ring receives a \text{\ttfamily Start} message,
it nominates itself to its successor in the ring.
\begin{hscode}\SaveRestoreHook
\column{B}{@{}>{\hspre}l<{\hspost}@{}}%
\column{3}{@{}>{\hspre}l<{\hspost}@{}}%
\column{5}{@{}>{\hspre}l<{\hspost}@{}}%
\column{E}{@{}>{\hspre}l<{\hspost}@{}}%
\>[B]{}\Varid{node}\;\Varid{state}\mathord{@}\Conid{Member}\;\!\{\Varid{next}\mskip1.5mu\}{}\<[E]%
\\
\>[B]{}\hsindent{3}{}\<[3]%
\>[3]{}\Conid{Envelope}\;\!\{\Varid{message}\mathrel{=}\Conid{Start}\mskip1.5mu\}\mathrel{=}\mathbf{do}{}\<[E]%
\\
\>[3]{}\hsindent{2}{}\<[5]%
\>[5]{}\Varid{self}\leftarrow \Varid{myThreadId}{}\<[E]%
\\
\>[3]{}\hsindent{2}{}\<[5]%
\>[5]{}\Varid{send}\;\Varid{next}\mathbin{\$}\Conid{Nominate}\;\Varid{self}{}\<[E]%
\\
\>[3]{}\hsindent{2}{}\<[5]%
\>[5]{}\Varid{return}\;\Varid{state}{}\<[E]%
\ColumnHook
\end{hscode}\resethooks
When a member of the ring receives a \text{\ttfamily Nominate} message, it compares the
nominee to its own identity.
If they are equal, then the member wins and the algorithm stops.
If the nominee is greater, then the member forwards the nomination to its
successor.
\begin{hscode}\SaveRestoreHook
\column{B}{@{}>{\hspre}l<{\hspost}@{}}%
\column{3}{@{}>{\hspre}l<{\hspost}@{}}%
\column{5}{@{}>{\hspre}l<{\hspost}@{}}%
\column{6}{@{}>{\hspre}l<{\hspost}@{}}%
\column{9}{@{}>{\hspre}c<{\hspost}@{}}%
\column{9E}{@{}l@{}}%
\column{12}{@{}>{\hspre}l<{\hspost}@{}}%
\column{20}{@{}>{\hspre}l<{\hspost}@{}}%
\column{24}{@{}>{\hspre}l<{\hspost}@{}}%
\column{E}{@{}>{\hspre}l<{\hspost}@{}}%
\>[B]{}\Varid{node}\;\Varid{state}\mathord{@}\Conid{Member}\;\!\{\Varid{next}\mskip1.5mu\}{}\<[E]%
\\
\>[B]{}\hsindent{3}{}\<[3]%
\>[3]{}\Conid{Envelope}\;\!\{\Varid{message}\mathrel{=}\Conid{Nominate}\;\!\{\Varid{nominee}\mathrel{=}\Varid{nom}\mskip1.5mu\}\mskip1.5mu\}\mathrel{=}\mathbf{do}{}\<[E]%
\\
\>[3]{}\hsindent{2}{}\<[5]%
\>[5]{}\Varid{self}\leftarrow \Varid{myThreadId}{}\<[E]%
\\
\>[3]{}\hsindent{2}{}\<[5]%
\>[5]{}\mathbf{case}\;()\;\mathbf{of}{}\<[E]%
\\
\>[5]{}\hsindent{1}{}\<[6]%
\>[6]{}\anonymous {}\<[9]%
\>[9]{}\mid {}\<[9E]%
\>[12]{}\Varid{self}\equiv \Varid{nom}\to \Varid{putStrLn}\;(\Varid{show}\;\Varid{self}\mathbin{+\hspace{-0.2em}+}\text{\ttfamily \char34 :~I~win\char34}){}\<[E]%
\\
\>[9]{}\mid {}\<[9E]%
\>[12]{}\Varid{self}\mathbin{<}{}\<[20]%
\>[20]{}\Varid{nom}\to \Varid{send}\;\Varid{next}\;(\Conid{Nominate}\;\Varid{nom}){}\<[E]%
\\
\>[9]{}\mid {}\<[9E]%
\>[12]{}\Varid{otherwise}{}\<[24]%
\>[24]{}\to \Varid{putStrLn}\;\text{\ttfamily \char34 Ignored~nomination\char34}{}\<[E]%
\\
\>[3]{}\hsindent{2}{}\<[5]%
\>[5]{}\Varid{return}\;\Varid{state}{}\<[E]%
\ColumnHook
\end{hscode}\resethooks
\ignore{
\begin{hscode}\SaveRestoreHook
\column{B}{@{}>{\hspre}l<{\hspost}@{}}%
\column{E}{@{}>{\hspre}l<{\hspost}@{}}%
\>[B]{}\Varid{node}\;\anonymous \;\anonymous \mathrel{=}\Varid{error}\;\text{\ttfamily \char34 node:~unhandled\char34}{}\<[E]%
\ColumnHook
\end{hscode}\resethooks
}

\subsubsection{Election initialization}
\label{sec:main1-init}

The election initialization function
is implemented in \Cref{fig:ringElection}.
It takes the size of the ring
and an unevaluated \text{\ttfamily IO} action representing node behavior,
and then performs the following steps to start an election:\footnote{
    The implementation shown doesn't handle rings of size 0 or 1.
    Also, we do not show thread cleanup.
}
\begin{enumerate}[leftmargin=2em]
    \item Create actors (with asynchronous exceptions masked).

    \item Randomize the order of actor \text{\ttfamily ThreadId}s.\footnote{The implementation of \text{\ttfamily permute} is in \Cref{apx:permute-impl}.}

    \item Inform each actor of the \text{\ttfamily ThreadId} that follows it in the
    random order (its successor) with an \text{\ttfamily Init} message.

    \item Send each actor the \text{\ttfamily Start} message to kick things off.
\end{enumerate}
To call the election initialization function, we construct an \text{\ttfamily IO} action
by passing the node intent function and the initial node state to the actor
main loop from \Cref{fig:run}:
\ignore{
\begin{hscode}\SaveRestoreHook
\column{B}{@{}>{\hspre}l<{\hspost}@{}}%
\column{5}{@{}>{\hspre}l<{\hspost}@{}}%
\column{E}{@{}>{\hspre}l<{\hspost}@{}}%
\>[B]{}\Varid{main1}\mathbin{::}\Conid{Int}\to \Conid{IO}\;(){}\<[E]%
\\
\>[B]{}\Varid{main1}\;\Varid{count}\mathrel{=}\mathbf{do}{}\<[E]%
\\
\>[B]{}\hsindent{5}{}\<[5]%
\>[5]{}\Varid{ring}\leftarrow {}\<[E]%
\ColumnHook
\end{hscode}\resethooks
}
\begin{center}
\begin{hscode}\SaveRestoreHook
\column{B}{@{}>{\hspre}l<{\hspost}@{}}%
\column{9}{@{}>{\hspre}l<{\hspost}@{}}%
\column{E}{@{}>{\hspre}l<{\hspost}@{}}%
\>[9]{}\Varid{ringElection}\;\Varid{count}\mathbin{\$}\Varid{run}\;\Varid{node}\;\Conid{Uninitialized}{}\<[E]%
\ColumnHook
\end{hscode}\resethooks
\end{center}
\ignore{
\begin{hscode}\SaveRestoreHook
\column{B}{@{}>{\hspre}l<{\hspost}@{}}%
\column{5}{@{}>{\hspre}l<{\hspost}@{}}%
\column{E}{@{}>{\hspre}l<{\hspost}@{}}%
\>[5]{}\Varid{threadDelay}\;\mathrm{1000000}{}\<[E]%
\\
\>[5]{}\Varid{mapM\char95 }\;\Varid{killThread}\;\Varid{ring}{}\<[E]%
\ColumnHook
\end{hscode}\resethooks
}
An election execution trace appears in \Cref{fig:main1-trace}.

\begin{figure}
\raggedright
\begin{hscode}\SaveRestoreHook
\column{B}{@{}>{\hspre}l<{\hspost}@{}}%
\column{5}{@{}>{\hspre}l<{\hspost}@{}}%
\column{9}{@{}>{\hspre}l<{\hspost}@{}}%
\column{E}{@{}>{\hspre}l<{\hspost}@{}}%
\>[B]{}\Varid{ringElection}\mathbin{::}\Conid{Int}\to \Conid{IO}\;()\to \Conid{IO}\;[\Conid{ThreadId}\mskip1.5mu]{}\<[E]%
\\
\>[B]{}\Varid{ringElection}\;\Varid{n}\;\Varid{actor}\mathrel{=}\mathbf{do}{}\<[E]%
\\
\>[B]{}\hsindent{5}{}\<[5]%
\>[5]{}\Varid{nodes}\leftarrow \Varid{sequence}\mathbin{\circ}\Varid{replicate}\;\Varid{n}\mathbin{\circ}\Varid{mask\char95 }\mathbin{\$}\Varid{forkIO}\;\Varid{actor}\quad\hfill (1){}\<[E]%
\\
\>[B]{}\hsindent{5}{}\<[5]%
\>[5]{}\Varid{ring}\leftarrow \Varid{getStdRandom}\mathbin{\$}\Varid{permute}\;\Varid{nodes}\quad\quad\hfill (2){}\<[E]%
\\
\>[B]{}\hsindent{5}{}\<[5]%
\>[5]{}\Varid{mapM\char95 }\;{}\<[E]%
\\
\>[5]{}\hsindent{4}{}\<[9]%
\>[9]{}(\lambda (\Varid{t},\Varid{next})\to \Varid{send}\;\Varid{t}\;\Conid{Init}\;\!\{\Varid{next}\mskip1.5mu\})\;\quad\quad\hfill (3){}\<[E]%
\\
\>[5]{}\hsindent{4}{}\<[9]%
\>[9]{}(\Varid{zip}\;\Varid{ring}\mathbin{\$}\Varid{tail}\;\Varid{ring}\mathbin{+\hspace{-0.2em}+}[\Varid{head}\;\Varid{ring}\mskip1.5mu]){}\<[E]%
\\
\>[B]{}\hsindent{5}{}\<[5]%
\>[5]{}\Varid{mapM\char95 }\;(\lambda \Varid{t}\to \Varid{send}\;\Varid{t}\;\Conid{Start})\;\Varid{ring}\quad\quad\hfill (4){}\<[E]%
\\
\>[B]{}\hsindent{5}{}\<[5]%
\>[5]{}\Varid{return}\;\Varid{ring}{}\<[E]%
\ColumnHook
\end{hscode}\resethooks
\caption{Ring leader election initialization.}
\label{fig:ringElection}
\end{figure}

\begin{figure}
\raggedright
\scriptsize

\begin{tabbing}\ttfamily
~ThreadId~46~send~Init~\char123{}next~\char61{}~ThreadId~50\char125{}~to~ThreadId~49\\
\ttfamily ~ThreadId~46~send~Init~\char123{}next~\char61{}~ThreadId~48\char125{}~to~ThreadId~50\\
\ttfamily ~ThreadId~46~send~Init~\char123{}next~\char61{}~ThreadId~47\char125{}~to~ThreadId~48\\
\ttfamily ~ThreadId~46~send~Init~\char123{}next~\char61{}~ThreadId~49\char125{}~to~ThreadId~47\\
\ttfamily ~ThreadId~46~send~Start~to~ThreadId~49\\
\ttfamily ~ThreadId~49~send~Nominate~\char123{}nominee~\char61{}~ThreadId~49\char125{}~to~ThreadId~50\\
\ttfamily ~ThreadId~46~send~Start~to~ThreadId~50\\
\ttfamily ~Ignored~nomination\\
\ttfamily ~ThreadId~50~send~Nominate~\char123{}nominee~\char61{}~ThreadId~50\char125{}~to~ThreadId~48\\
\ttfamily ~ThreadId~46~send~Start~to~ThreadId~48\\
\ttfamily ~ThreadId~48~send~Nominate~\char123{}nominee~\char61{}~ThreadId~50\char125{}~to~ThreadId~47\\
\ttfamily ~ThreadId~47~send~Nominate~\char123{}nominee~\char61{}~ThreadId~50\char125{}~to~ThreadId~49\\
\ttfamily ~ThreadId~48~send~Nominate~\char123{}nominee~\char61{}~ThreadId~48\char125{}~to~ThreadId~47\\
\ttfamily ~ThreadId~46~send~Start~to~ThreadId~47\\
\ttfamily ~ThreadId~49~send~Nominate~\char123{}nominee~\char61{}~ThreadId~50\char125{}~to~ThreadId~50\\
\ttfamily ~ThreadId~47~send~Nominate~\char123{}nominee~\char61{}~ThreadId~48\char125{}~to~ThreadId~49\\
\ttfamily ~ThreadId~50\char58{}~I~win\\
\ttfamily ~Ignored~nomination\\
\ttfamily ~ThreadId~47~send~Nominate~\char123{}nominee~\char61{}~ThreadId~47\char125{}~to~ThreadId~49\\
\ttfamily ~Ignored~nomination
\end{tabbing}

\normalsize
\caption{An execution trace of the ring leader election solution.}
\label{fig:main1-trace}
\end{figure}

\subsection{Extending the leader election}
\label{sec:dyn-ring}

The solution we have shown solves the ring leader election problem
insofar as a single node concludes that it has won.
However, it is also desirable for the other nodes to learn the outcome of the
election.
Since it is sometimes necessary to extend a system without modifying the
original, we will show how to extend the original ring leader election to add a
winner-declaration round.

Since there is no message constructor to inform nodes of the election outcome,
we will define a new message type whose constructor indicates a declaration of
who is the winner.
We will extend the existing node intent function by wrapping it with a new
intent function that processes messages of either the old or the new message
types, with distinct behavior for each, leveraging the dynamic types support
described in \Cref{sec:dynamic-types}.
The new behaviors are:
\begin{itemize}[leftmargin=1.5em]
    \item[--] Each node remembers the greatest nominee it has seen.

    \item[--] When the winner self-identifies, they will start an extra round
    declaring themselves winner.

    \item[--] Upon receiving a winner declaration, a node compares the greatest
    nominee it has seen with the declared-winner.
    If they are the same, then the node forwards the declaration to its
    successor.
\end{itemize}

Extended nodes will store the original node state (\Cref{fig:node-types})
paired with the identity of the greatest nominee they have seen.
This new extended node state is shown in \Cref{fig:exnode-types} as type
\text{\ttfamily Exnode}.
The new message type (\text{\ttfamily Winner}, also in \Cref{fig:exnode-types}) has only
one constructor and is used to declare some node the winner.

\begin{figure}
\raggedright
\begin{hscode}\SaveRestoreHook
\column{B}{@{}>{\hspre}l<{\hspost}@{}}%
\column{E}{@{}>{\hspre}l<{\hspost}@{}}%
\>[B]{}\mathbf{type}\;\Conid{Exnode}\mathrel{=}(\Conid{Node},\Conid{ThreadId}){}\<[E]%
\\[\blanklineskip]%
\>[B]{}\mathbf{data}\;\Conid{Winner}\mathrel{=}\Conid{Winner}\;\Conid{ThreadId}\;\mathbf{deriving}\;\Conid{Show}{}\<[E]%
\\[\blanklineskip]%
\>[B]{}\mathbf{instance}\;\Conid{Exception}\;\Conid{Winner}{}\<[E]%
\ColumnHook
\end{hscode}\resethooks
\caption{
    Extended nodes store node state alongside the greatest nominee
    seen.
    They accept one message in addition to those in \Cref{fig:node-types}.
}
\label{fig:exnode-types}
\end{figure}

\subsubsection{Declaration-round termination}
When an extended node receives a declaration of the winner
that matches their greatest nominee seen,
they have ``learned'' that that node is indeed the winner.
When the winner receives their own declaration,
\emph{everyone} has learned they are the winner,
and the algorithm terminates.

\subsubsection{Exnode-actor behavior}
\label{sec:ring2-intent-fun}

The intent function for the new actor will have state \text{\ttfamily Exnode} and
receive messages of type \text{\ttfamily SomeException}.
This will allow it to receive either \text{\ttfamily Msg} or \text{\ttfamily Winner} values and
branch on which is received.
\begin{hscode}\SaveRestoreHook
\column{B}{@{}>{\hspre}l<{\hspost}@{}}%
\column{E}{@{}>{\hspre}l<{\hspost}@{}}%
\>[B]{}\Varid{exnode}\mathbin{::}\Conid{Intent}\;\Conid{Exnode}\;\Conid{SomeException}{}\<[E]%
\ColumnHook
\end{hscode}\resethooks

Recall the implementation of the actor main loop function,
\text{\ttfamily runDyn} from \Cref{fig:dyn-impl}.
When we apply \text{\ttfamily exnode} to \text{\ttfamily runDyn},
the call to \text{\ttfamily fromException} in \text{\ttfamily runDyn}
is inferred to return \text{\ttfamily Maybe~SomeException},
which succeeds unconditionally.
The \text{\ttfamily exnode} intent function must then perform its own downcasts,
and we enable \text{\ttfamily ViewPatterns} to ease our presentation.
There are two main cases,
corresponding to the two message types the actor will handle,
which we explain below.

The first case of \text{\ttfamily exnode}, shown in \Cref{fig:exnode-case-msg}, applies
when an extended node downcasts the envelope contents to \text{\ttfamily Msg}.
In each of its branches, node state is updated by delegating part of message
handling to the held node.
We annotate the rest of \Cref{fig:exnode-case-msg} as follows:
\begin{enumerate}[leftmargin=2em]
    \item Delegate to the held node by putting the revealed \text{\ttfamily Msg} back
    into its envelope and passing it through the intent function, \text{\ttfamily node},
    from \Cref{sec:ring-intent-fun}.
    \item If the message is a nomination of the current node, start
    the winner round, because the election is over.
    \item Otherwise, the election is ongoing, so keep track of the greatest
    nominee seen.
\end{enumerate}
\begin{figure}
\raggedright
\begin{hscode}\SaveRestoreHook
\column{B}{@{}>{\hspre}l<{\hspost}@{}}%
\column{3}{@{}>{\hspre}l<{\hspost}@{}}%
\column{5}{@{}>{\hspre}l<{\hspost}@{}}%
\column{9}{@{}>{\hspre}l<{\hspost}@{}}%
\column{13}{@{}>{\hspre}l<{\hspost}@{}}%
\column{17}{@{}>{\hspre}l<{\hspost}@{}}%
\column{E}{@{}>{\hspre}l<{\hspost}@{}}%
\>[B]{}\Varid{exnode}\;(\Varid{n},\Varid{great})\;{}\<[E]%
\\
\>[B]{}\hsindent{3}{}\<[3]%
\>[3]{}\Varid{e}\mathord{@}\Conid{Envelope}\;\!\{\Varid{message}\mathrel{=}\Varid{fromException}\to \Conid{Just}\;\Varid{m}\mskip1.5mu\}\mathrel{=}\mathbf{do}{}\<[E]%
\\
\>[3]{}\hsindent{2}{}\<[5]%
\>[5]{}\Varid{self}\leftarrow \Varid{myThreadId}{}\<[E]%
\\
\>[3]{}\hsindent{2}{}\<[5]%
\>[5]{}\Varid{n'}\mathord{@}\Conid{Member}\;\!\{\Varid{next}\mskip1.5mu\}\leftarrow \Varid{node}\;\Varid{n}\;\Varid{e}\;\!\{\Varid{message}\mathrel{=}\Varid{m}\mskip1.5mu\}\quad\quad\hfill (1){}\<[E]%
\\
\>[3]{}\hsindent{2}{}\<[5]%
\>[5]{}\mathbf{case}\;\Varid{m}\;\mathbf{of}{}\<[E]%
\\
\>[5]{}\hsindent{4}{}\<[9]%
\>[9]{}\Conid{Nominate}\;\!\{\Varid{nominee}\mskip1.5mu\}\to {}\<[E]%
\\
\>[9]{}\hsindent{4}{}\<[13]%
\>[13]{}\mathbf{if}\;\Varid{self}\equiv \Varid{nominee}{}\<[E]%
\\
\>[9]{}\hsindent{4}{}\<[13]%
\>[13]{}\mathbf{then}\;\Varid{send}\;\Varid{next}\;(\Conid{Winner}\;\Varid{self})\quad\quad\hfill (2){}\<[E]%
\\
\>[13]{}\hsindent{4}{}\<[17]%
\>[17]{}\mathbin{>\hspace{-0.4em}>}\Varid{return}\;(\Varid{n'},\Varid{great}){}\<[E]%
\\
\>[9]{}\hsindent{4}{}\<[13]%
\>[13]{}\mathbf{else}\;\Varid{return}\;(\Varid{n'},\Varid{max}\;\Varid{nominee}\;\Varid{great})\quad\quad\hfill (3){}\<[E]%
\\
\>[5]{}\hsindent{4}{}\<[9]%
\>[9]{}\anonymous \to \Varid{return}\;(\Varid{n'},\Varid{great}){}\<[E]%
\ColumnHook
\end{hscode}\resethooks
\caption{
    When \text{\ttfamily exnode} receives a \text{\ttfamily Msg}, it delegates to \text{\ttfamily node}.
    It may also update the greatest nominee seen or trigger the
    winner-declaration round.
}
\label{fig:exnode-case-msg}
\end{figure}

The second case of \text{\ttfamily exnode} applies when a node downcasts the envelope
contents to a winner declaration.
Its implementation is shown in \Cref{fig:exnode-case-winner}.
If the current node is declared winner, the algorithm terminates successfully.
If the greatest nominee the current node has seen is declared winner, the node
forwards the declaration to its successor.
State is unchanged in each of these branches.
\begin{figure}
\raggedright
\begin{hscode}\SaveRestoreHook
\column{B}{@{}>{\hspre}l<{\hspost}@{}}%
\column{3}{@{}>{\hspre}l<{\hspost}@{}}%
\column{5}{@{}>{\hspre}l<{\hspost}@{}}%
\column{9}{@{}>{\hspre}l<{\hspost}@{}}%
\column{13}{@{}>{\hspre}l<{\hspost}@{}}%
\column{E}{@{}>{\hspre}l<{\hspost}@{}}%
\>[B]{}\Varid{exnode}\;\Varid{state}\mathord{@}(\Conid{Member}\;\!\{\Varid{next}\mskip1.5mu\},\Varid{great}){}\<[E]%
\\
\>[B]{}\hsindent{3}{}\<[3]%
\>[3]{}\Conid{Envelope}\;\!\{\Varid{message}\mathrel{=}\Varid{fromException}\to \Conid{Just}\;\Varid{m}\mskip1.5mu\}\mathrel{=}\mathbf{do}{}\<[E]%
\\
\>[3]{}\hsindent{2}{}\<[5]%
\>[5]{}\Varid{self}\leftarrow \Varid{myThreadId}{}\<[E]%
\\
\>[3]{}\hsindent{2}{}\<[5]%
\>[5]{}\mathbf{case}\;\Varid{m}\;\mathbf{of}{}\<[E]%
\\
\>[5]{}\hsindent{4}{}\<[9]%
\>[9]{}\Conid{Winner}\;\Varid{w}{}\<[E]%
\\
\>[9]{}\hsindent{4}{}\<[13]%
\>[13]{}\mid \Varid{w}\equiv \Varid{self}\to \Varid{putStrLn}\;(\Varid{show}\;\Varid{self}\mathbin{+\hspace{-0.2em}+}\text{\ttfamily \char34 :~Confirmed\char34}){}\<[E]%
\\
\>[9]{}\hsindent{4}{}\<[13]%
\>[13]{}\mid \Varid{w}\equiv \Varid{great}\to \Varid{send}\;\Varid{next}\;(\Conid{Winner}\;\Varid{w}){}\<[E]%
\\
\>[9]{}\hsindent{4}{}\<[13]%
\>[13]{}\mid \Varid{otherwise}\to \Varid{putStrLn}\;\text{\ttfamily \char34 Unexpected~winner\char34}{}\<[E]%
\\
\>[3]{}\hsindent{2}{}\<[5]%
\>[5]{}\Varid{return}\;\Varid{state}{}\<[E]%
\ColumnHook
\end{hscode}\resethooks
\caption{
    When \text{\ttfamily exnode} receives a \text{\ttfamily Winner}, it manages the
    winner-declaration round.
}
\label{fig:exnode-case-winner}
\end{figure}

\ignore{
\begin{hscode}\SaveRestoreHook
\column{B}{@{}>{\hspre}l<{\hspost}@{}}%
\column{E}{@{}>{\hspre}l<{\hspost}@{}}%
\>[B]{}\Varid{exnode}\;\anonymous \;\anonymous \mathrel{=}\Varid{error}\;\text{\ttfamily \char34 exnode:~unhandled\char34}{}\<[E]%
\ColumnHook
\end{hscode}\resethooks
}

\subsubsection{Extended election initialization}
\label{sec:main2-init}

The extended ring leader election reuses the
initialization scaffolding from before
(\Cref{fig:ringElection}).
The only change is that the \text{\ttfamily IO} action passed to
\text{\ttfamily ringElection} initializes the greatest nominee seen
to itself, prior to calling \text{\ttfamily run}.
It is called like this:
\ignore{
\begin{hscode}\SaveRestoreHook
\column{B}{@{}>{\hspre}l<{\hspost}@{}}%
\column{5}{@{}>{\hspre}l<{\hspost}@{}}%
\column{E}{@{}>{\hspre}l<{\hspost}@{}}%
\>[B]{}\Varid{main2}\mathbin{::}\Conid{Int}\to \Conid{IO}\;(){}\<[E]%
\\
\>[B]{}\Varid{main2}\;\Varid{count}\mathrel{=}\mathbf{do}{}\<[E]%
\\
\>[B]{}\hsindent{5}{}\<[5]%
\>[5]{}\Varid{ring}\leftarrow {}\<[E]%
\ColumnHook
\end{hscode}\resethooks
}
\begin{center}
\begin{hscode}\SaveRestoreHook
\column{B}{@{}>{\hspre}l<{\hspost}@{}}%
\column{9}{@{}>{\hspre}l<{\hspost}@{}}%
\column{13}{@{}>{\hspre}l<{\hspost}@{}}%
\column{E}{@{}>{\hspre}l<{\hspost}@{}}%
\>[9]{}\Varid{ringElection}\;\Varid{count}\mathbin{\$}\mathbf{do}{}\<[E]%
\\
\>[9]{}\hsindent{4}{}\<[13]%
\>[13]{}\Varid{great}\leftarrow \Varid{myThreadId}{}\<[E]%
\\
\>[9]{}\hsindent{4}{}\<[13]%
\>[13]{}\Varid{run}\;\Varid{exnode}\;(\Conid{Uninitialized},\Varid{great}){}\<[E]%
\ColumnHook
\end{hscode}\resethooks
\end{center}
\ignore{
\begin{hscode}\SaveRestoreHook
\column{B}{@{}>{\hspre}l<{\hspost}@{}}%
\column{5}{@{}>{\hspre}l<{\hspost}@{}}%
\column{E}{@{}>{\hspre}l<{\hspost}@{}}%
\>[5]{}\Varid{threadDelay}\;\mathrm{1000000}{}\<[E]%
\\
\>[5]{}\Varid{mapM\char95 }\;\Varid{killThread}\;\Varid{ring}{}\<[E]%
\ColumnHook
\end{hscode}\resethooks
}
A trace of an extended election appears in \Cref{apx:main2-trace}.

\section{What have we wrought?}
\label{sec:what-have-we-wrought}

\Cref{fig:static-impl} shows that we have, in only a few lines of
code, discovered an actor framework within GHC's RTS that makes no explicit use
of channels, references, or locks and imports just a few names from default
modules.
The support for dynamic types, shown in \Cref{fig:dyn-impl} as separate
definitions, can be folded into \Cref{fig:static-impl} for only a few
additional lines.\footnote{
    Instead of wrapping the intent function, the framework's message downcast
    is performed in the exception handler.
}
We find it intriguing that this is possible and shocking that it is so easy.

\subsection{Almost a COPL}
\label{sec:almost-copl}

In \Cref{sec:actor-model} we described an actor framework as having the
characteristics of a \emph{concurrency-oriented programming language}
(COPL)~\citep{armstrong2003}.
Which of the COPL requirements does our framework satisfy?
Here we review the criteria listed in \Cref{sec:actor-model}:
%
%
\begin{enumerate}[leftmargin=2em]
    \item \ding{51} Threads behave as independent processes.
    \item \ding{55}/\ding{51} Threads are not strongly isolated because
    termination of the main thread terminates all others. However, if the main
    thread is excluded as a special case, then the set of other threads are
    strongly isolated.
    \item \ding{51} \text{\ttfamily ThreadID} is unique, hidden, and unforgeable.
    \item \ding{55} Threads may have shared state.
    \item \ding{55} Asynchronous exceptions do not behave as \emph{unreliable} message passing.
    \item \ding{51} An actor can reliably inform others when it halts using
    \text{\ttfamily forkFinally}.
\end{enumerate}

The message-passing semantics of our actor framework is nuanced.
Documentation for the interfaces we use indicates that the framework provides
\emph{reliable synchronous message passing with FIFO order}.
We call it \emph{synchronous} because ``\text{\ttfamily throwTo} does not return until
the exception is received by the target thread''
\cite{controlDotException}.\footnote{
    ``Synchronous for me, but not for thee'' might be the most correct
    characterization. Senders may experience GHC's asynchronous exceptions as
    synchronous, but recipients will always perceive them as asynchronous.
}
This means that a sender may block if the recipient is not well-behaved (e.g.,
its intent function enters an infinite loop in pure computation).
We distinguish \emph{well-behaved} intent functions, which eventually
terminate or reach an interruptible point,
from \emph{poorly-behaved} intent functions, which do not.
Assuming intent functions are well-behaved,
the framework will tend to exhibit the behavior of
\emph{reliable asynchronous message passing with FIFO order}
and occasional double-sends,
because senders will not observe the blocking behavior of \text{\ttfamily throwTo}.
By wrapping calls to the send function with \text{\ttfamily forkIO}~\cite{marlow2001async}, we can achieve
\emph{reliable asynchronous message passing without FIFO order}
even in the presence of poorly-behaved intent functions.\footnote{
    If thread $T_1$ forks thread $T_2$ to send message $M_2$, and then $T_1$
    forks thread $T_3$ to send message $M_3$, the RTS scheduler may first run
    $T_3$ resulting in $M_3$ reaching the recipient before $M_2$, violating
    FIFO if both messages have the same recipient.
}
FIFO can then be recovered by message sequence numbers or by (albeit, jumping the
shark) use of an outbox thread per actor.
With those caveats in mind, our framework \emph{mostly} satisfies \citet{armstrong2003}'s criteria for message-passing semantics:
%
%
\begin{enumerate}[leftmargin=2em]
    \item[(5a)] \ding{55}/\ding{51} A stuck recipient may cause a sender to become stuck,
    unless senders use \text{\ttfamily forkIO}
    or we assume the recipient is well-behaved.

    \item[(5b)] \ding{55}/\ding{51} Actors know that a message is \emph{received} (stored
    in the recipient inbox) as soon as \text{\ttfamily send} returns.
    However, they do not know that a message is \emph{delivered} (processed by
    the recipient) until receiving a response.

    \item[(5c)] \ding{51}/\ding{55} Messages between two actors obey FIFO ordering,
    unless \text{\ttfamily forkIO} is used when sending.
\end{enumerate}

Our choice to wrap a user-defined message type in a known envelope type has the
benefit of allowing the actor main loop to distinguish between messages and
exceptions, allowing the latter to terminate the thread as intended.
At the same time, though, this choice runs afoul of the \emph{name distribution problem}~\cite{armstrong2003} by indiscriminately informing all recipients of the sender
process identifier.
One strategy to hide to an actor's name and restore the lost security isolation is
to wrap calls to the send function with \text{\ttfamily forkIO}.
Another strategy would be to define two constructors for envelope, and elide
the ``sender'' field from one.

We claim that our actor framework is \emph{almost} a COPL.
It also meets our informal requirements that actors can send and receive
messages, update state, and spawn or kill other actors (though we have not
shown examples of all of these).
However, we do not mean to imply that our actor framework is practical; we
merely mean to point out that it is, indeed, an actor framework.

\subsection{Summary of performance evaluation}

We have described a novel approach to inter-thread communication.
We believe it
is prudent to compare the performance this \emph{unintended communication
mechanism} against the performance of an \emph{intended communication
mechanism} to restore a sense that the ship is indeed upright.
To that end,
we re-implemented the extended ring leader election from \Cref{sec:ring-impl}
using channels --- a standard FIFO communication primitive.
We also implemented a ``control''\footnote{
    The ``control'' forks some number of threads that do nothing and
    immediately kills them.
} to establish a lower bound on the expected
running time of the actor-based and channel-based implementations.

%
We compared the running time of these implementations
at ring sizes up to $65536$ nodes on machines with 8, 32, and 192
capabilities.
We also compared their total allocations over the program run at various ring
sizes.

\begin{figure}
\raggedright

    \begin{subfigure}{\linewidth}
        \begin{small}
        \def\svgwidth{\linewidth}
        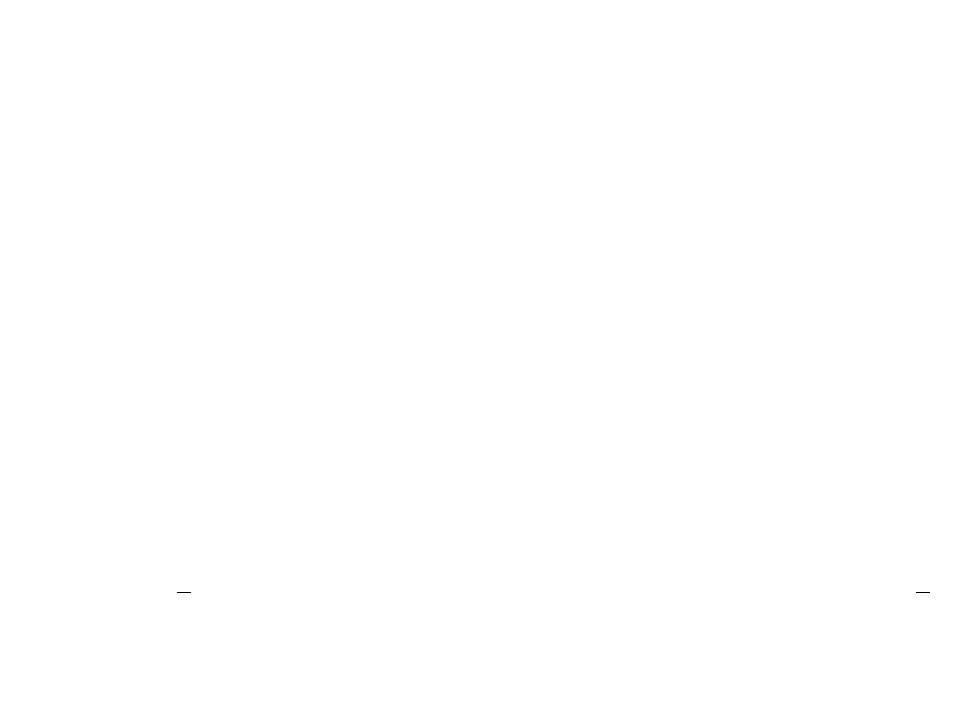
        \end{small}
        \caption{
            The channel-based implementation is faster than the actor-based
            implementation, except at very large numbers of threads.
            We reproduced this result on machines with 8, 32, and 192 capabilities.
        }
        \label{fig:perf-eval-time-n32}
    \end{subfigure}

    \begin{subfigure}{\linewidth}
        \begin{small}
        \def\svgwidth{\linewidth}
        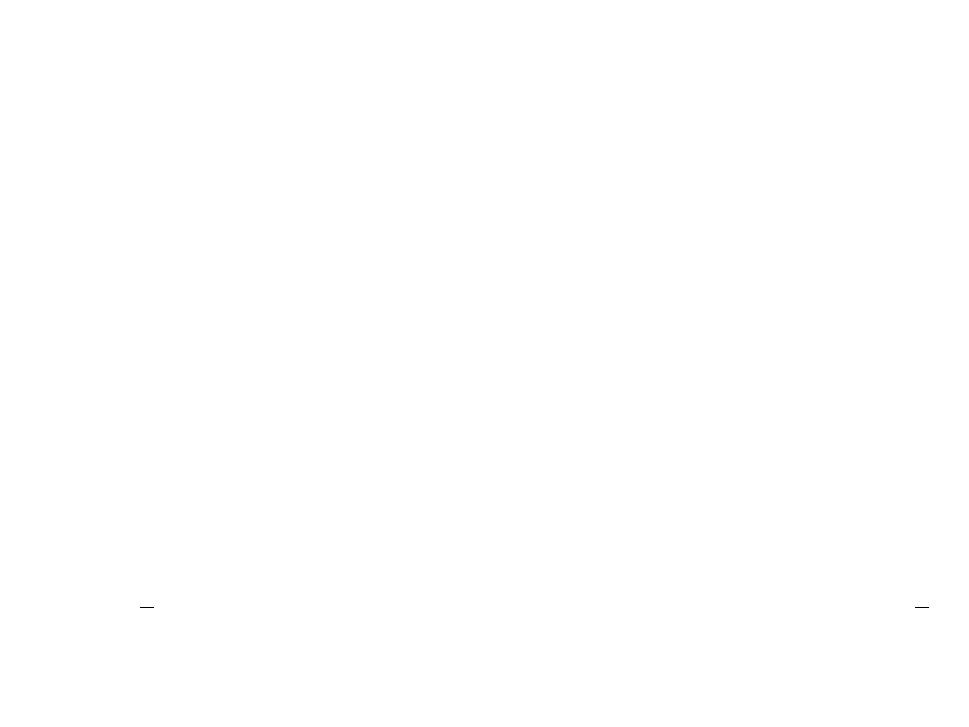
        \end{small}
        \caption{
            The growth of allocations by the channel-based implementation
            eventually catches up to that of the actor-based implementation.
        }
        \label{fig:perf-eval-mem}
    \end{subfigure}

\caption{Representative selection of experimental results.}
\label{fig:perf-eval}
\end{figure}

%
Our running time results (\Cref{fig:perf-eval-time-n32}) show that
the actor-based implementation is significantly slower
than the channel-based implementation for ring sizes less than $8192$ nodes,
but surprisingly, it is marginally faster for more than $32768$ nodes.
The total-allocations result (\Cref{fig:perf-eval-mem}) shows that
allocations made by the channel-based implementation
catch up to that of the actor-based implementation at large ring sizes,
and we hypothesize that this convergence
explains why the running time results swap places.
Additionally, our results show that the running time of the extended ring leader
election algorithm is invariant to the number of capabilities used by the RTS,
making it a poor choice for a general evaluation of our actor framework,
but sufficient for our purpose of confirming that channels are faster.

\Cref{apx:actor-bench-impl,apx:control-bench-impl,apx:channel-bench-impl,apx:criterion-bench-impl}
give the source code for these benchmarks.
\Cref{apx:exp-setup} details our experimental setup, and
\Cref{apx:exp-result} discusses more of the results.

\section{Conclusion}
\label{sec:conclusion}

Can we implement an actor framework with Haskell's threads and asynchronous
exceptions?
Our implementation and results show that we can, and this fact hints that
perhaps asynchronous exceptions are at least as general as actors.

However, the actor framework we present is not an advancement:
It is easy to use, but easy to use wrongly.
It has acceptable throughput, but is slower than accepted
tools.
It requires no appreciable dependencies, no explicitly mutable data structures
or references, no effort to achieve synchronization, and very little code only
because \emph{those things already exist, abstracted within the RTS}.

\emph{Should} it have been possible to implement the actor framework we present
here?
Like many people, we choose Haskell because it is a tool that typically
prevents ``whole classes of errors,'' and also because it is a joy to use.
But in this paper we achieve dynamically typed ``spooky action at a distance''
with frighteningly little effort.
Perhaps the user-accessible interface to the asynchronous exception system
should be constrained.

More broadly,
with the 9.6.1 release of  GHC,
a user of the RTS enjoys
software transactional memory,
asynchronous exceptions,
delimited continuations (and extensible algebraic effects),
and more,
together in the same tub.
The water is warm --- jump in!
Will all the members of this new \emph{extended} ``awkward squad''
\cite{peytonjones2001tackling} bob gently together, or will they knock elbows?
Which of them can be implemented in terms of the
others, and should their full power be exposed so that we can do so?
We hope the reader will draw their own conclusions.

\begin{acks}
This paper grew out of a presentation at Portland State University's Programming Languages \& Verification group in May 2023, and we are grateful to the Portland State PLV group for their enthusiasm and encouragement of our work.
We also thank Jos\'{e} Calder\'{o}n, the members of the LSD Lab at UC Santa Cruz, and the anonymous Haskell '23 reviewers for their valuable feedback on drafts of this paper.

This material is based upon work supported by the National Science Foundation under Grant No. CCF-2145367. Any opinions, findings, and conclusions or recommendations expressed in this material are those of the author(s) and do not necessarily reflect the views of the National Science Foundation.
\end{acks}

\bibliographystyle{ACM-Reference-Format}
\bibliography{main.bib}

\appendix

\section{Appendix}

\subsection{Permute function implementation}
\label{apx:permute-impl}

In \Cref{sec:ring-impl} we provided the implementation of a ring
leader election in our actor framework.
The implementation used \text{\ttfamily permute} to randomize the list of
\text{\ttfamily ThreadId}.
The \text{\ttfamily permute} function repeatedly pops a random element from the input and adds it to the output.
Its implementation is as follows:
\begin{hscode}\SaveRestoreHook
\column{B}{@{}>{\hspre}l<{\hspost}@{}}%
\column{3}{@{}>{\hspre}l<{\hspost}@{}}%
\column{5}{@{}>{\hspre}l<{\hspost}@{}}%
\column{9}{@{}>{\hspre}l<{\hspost}@{}}%
\column{13}{@{}>{\hspre}l<{\hspost}@{}}%
\column{E}{@{}>{\hspre}l<{\hspost}@{}}%
\>[B]{}\Varid{permute}\mathbin{::}\Conid{RandomGen}\;\Varid{g}\Rightarrow [\Varid{a}\mskip1.5mu]\to \Varid{g}\to ([\Varid{a}\mskip1.5mu],\Varid{g}){}\<[E]%
\\
\>[B]{}\Varid{permute}\;\Varid{pool0}\;\Varid{gen0}{}\<[E]%
\\
\>[B]{}\hsindent{5}{}\<[5]%
\>[5]{}\mathrel{=}\Varid{snd}{}\<[E]%
\\
\>[B]{}\hsindent{5}{}\<[5]%
\>[5]{}\mathbin{\circ}\Varid{foldr}\;\Varid{pick}\;(\Varid{pool0},([\mskip1.5mu],\Varid{gen0})){}\<[E]%
\\
\>[B]{}\hsindent{5}{}\<[5]%
\>[5]{}\mathbin{\$}\Varid{replicate}\;(\Varid{length}\;\Varid{pool0})\;(){}\<[E]%
\\
\>[B]{}\hsindent{3}{}\<[3]%
\>[3]{}\mathbf{where}{}\<[E]%
\\
\>[3]{}\hsindent{2}{}\<[5]%
\>[5]{}\Varid{pick}\;()\;(\Varid{pool},(\Varid{output},\Varid{g}))\mathrel{=}{}\<[E]%
\\
\>[5]{}\hsindent{4}{}\<[9]%
\>[9]{}\mathbf{let}\;(\Varid{index},\Varid{g'})\mathrel{=}\Varid{randomR}\;(\mathrm{0},\Varid{length}\;\Varid{pool}\mathbin{-}\mathrm{1})\;\Varid{g}{}\<[E]%
\\
\>[9]{}\hsindent{4}{}\<[13]%
\>[13]{}(\Varid{x},\Varid{pool'})\mathrel{=}\Varid{pop}\;\Varid{pool}\;\Varid{index}{}\<[E]%
\\
\>[5]{}\hsindent{4}{}\<[9]%
\>[9]{}\mathbf{in}\;(\Varid{pool'},(\Varid{x}\mathbin{:}\Varid{output},\Varid{g'})){}\<[E]%
\\
\>[3]{}\hsindent{2}{}\<[5]%
\>[5]{}\Varid{pop}\;(\Varid{x}\mathbin{:}\Varid{xs})\;\mathrm{0}\mathrel{=}(\Varid{x},\Varid{xs}){}\<[E]%
\\
\>[3]{}\hsindent{2}{}\<[5]%
\>[5]{}\Varid{pop}\;(\Varid{x}\mathbin{:}\Varid{xs})\;\Varid{n}\mathrel{=}(\Varid{x}\mathbin{:})\mathbin{\langle\$\rangle}\Varid{pop}\;\Varid{xs}\;(\Varid{n}\mathbin{-}\mathrm{1}){}\<[E]%
\\
\>[3]{}\hsindent{2}{}\<[5]%
\>[5]{}\Varid{pop}\;[\mskip1.5mu]\;\anonymous \mathrel{=}\Varid{error}\;\text{\ttfamily \char34 pop~empty~list\char34}{}\<[E]%
\ColumnHook
\end{hscode}\resethooks

\subsection{Actor benchmark implementation}
\label{apx:actor-bench-impl}

In the extended ring leader election solution, the time to
termination is
the time necessary for the winner's self-nomination to pass around the ring
once, plus the time for the winner-declaration to pass around the ring once.
Termination is detected when a node receives its own winner declaration.

We extend \text{\ttfamily exnode} (\Cref{sec:ring2-intent-fun})
to make a benchmark-node
with additional behavior:
When a benchmark-node is confirmed as winner, it puts its own
\text{\ttfamily ThreadId} into an \text{\ttfamily MVar} to signal termination.
\begin{hscode}\SaveRestoreHook
\column{B}{@{}>{\hspre}l<{\hspost}@{}}%
\column{5}{@{}>{\hspre}l<{\hspost}@{}}%
\column{9}{@{}>{\hspre}l<{\hspost}@{}}%
\column{E}{@{}>{\hspre}l<{\hspost}@{}}%
\>[B]{}\Varid{benchNode}\mathbin{::}\Conid{\Conid{Mv}.MVar}\;\Conid{ThreadId}\to \Conid{Intent}\;\Conid{Exnode}\;\Conid{SomeException}{}\<[E]%
\\
\>[B]{}\Varid{benchNode}\;\Varid{done}\;\Varid{state}\;\Varid{e}\mathord{@}\Conid{Envelope}\;\!\{\Varid{message}\mskip1.5mu\}\mathrel{=}\mathbf{do}{}\<[E]%
\\
\>[B]{}\hsindent{5}{}\<[5]%
\>[5]{}\Varid{state'}\leftarrow \Varid{exnode}\;\Varid{state}\;\Varid{e}{}\<[E]%
\\
\>[B]{}\hsindent{5}{}\<[5]%
\>[5]{}\Varid{self}\leftarrow \Varid{myThreadId}{}\<[E]%
\\
\>[B]{}\hsindent{5}{}\<[5]%
\>[5]{}\mathbf{case}\;\Varid{fromException}\;\Varid{message}\;\mathbf{of}{}\<[E]%
\\
\>[5]{}\hsindent{4}{}\<[9]%
\>[9]{}\Conid{Just}\;(\Conid{Winner}\;\Varid{w})\mid \Varid{w}\equiv \Varid{self}\to \Varid{\Conid{Mv}.putMVar}\;\Varid{done}\;\Varid{w}{}\<[E]%
\\
\>[5]{}\hsindent{4}{}\<[9]%
\>[9]{}\anonymous \to \Varid{return}\;(){}\<[E]%
\\
\>[B]{}\hsindent{5}{}\<[5]%
\>[5]{}\Varid{return}\;\Varid{state'}{}\<[E]%
\ColumnHook
\end{hscode}\resethooks

The reason we aren't using message passing to
notify about termination is because it is difficult to communicate between the
``actor world'' and the ``functional world.''
The functional world expects \text{\ttfamily IO} actions to terminate with return values,
but we didn't bother to implement clean termination in our actor framework.
Lacking that, we could try spawning an actor and then setting up an
exception handler to receive messages from it, but we choose not to do this
because of the potential for race conditions.

We benchmark time to termination using the \text{\ttfamily criterion} package.
For this, we need an \text{\ttfamily IO} action that executes the algorithm, cleans
up its resources, and then returns.
The function \text{\ttfamily benchActors} does this:
it runs an election with benchmark-nodes, waits for termination, kills the
nodes, and asserts a correct result.
\begin{hscode}\SaveRestoreHook
\column{B}{@{}>{\hspre}l<{\hspost}@{}}%
\column{5}{@{}>{\hspre}l<{\hspost}@{}}%
\column{9}{@{}>{\hspre}l<{\hspost}@{}}%
\column{E}{@{}>{\hspre}l<{\hspost}@{}}%
\>[B]{}\Varid{benchActors}\mathbin{::}\Conid{Int}\to \Conid{IO}\;(){}\<[E]%
\\
\>[B]{}\Varid{benchActors}\;\Varid{n}\mathrel{=}\mathbf{do}{}\<[E]%
\\
\>[B]{}\hsindent{5}{}\<[5]%
\>[5]{}\mbox{\onelinecomment  Start the ring-leader election}{}\<[E]%
\\
\>[B]{}\hsindent{5}{}\<[5]%
\>[5]{}\Varid{done}\leftarrow \Varid{\Conid{Mv}.newEmptyMVar}{}\<[E]%
\\
\>[B]{}\hsindent{5}{}\<[5]%
\>[5]{}\Varid{ring}\leftarrow \Varid{ringElection}\;\Varid{n}\mathbin{\$}\mathbf{do}{}\<[E]%
\\
\>[5]{}\hsindent{4}{}\<[9]%
\>[9]{}\Varid{great}\leftarrow \Varid{myThreadId}{}\<[E]%
\\
\>[5]{}\hsindent{4}{}\<[9]%
\>[9]{}\Varid{run}\;(\Varid{benchNode}\;\Varid{done})\;(\Conid{Uninitialized},\Varid{great}){}\<[E]%
\\
\>[B]{}\hsindent{5}{}\<[5]%
\>[5]{}\mbox{\onelinecomment  Wait for termination, kill the ring, assert correct result}{}\<[E]%
\\
\>[B]{}\hsindent{5}{}\<[5]%
\>[5]{}\Varid{w}\leftarrow \Varid{\Conid{Mv}.takeMVar}\;\Varid{done}{}\<[E]%
\\
\>[B]{}\hsindent{5}{}\<[5]%
\>[5]{}\Varid{mapM\char95 }\;\Varid{killThread}\;\Varid{ring}{}\<[E]%
\\
\>[B]{}\hsindent{5}{}\<[5]%
\>[5]{}\Varid{assert}\;(\Varid{w}\equiv \Varid{maximum}\;\Varid{ring})\;(\Varid{return}\;()){}\<[E]%
\ColumnHook
\end{hscode}\resethooks

\subsection{Control benchmark implementation}
\label{apx:control-bench-impl}

The experimental control, \text{\ttfamily benchControl}, only forks threads and then
kills them.
It is useful to establish whether or not, for example, laziness has caused
our non-control implementations to perform no work.
The other implementations should take longer than the control because they
do more work.
\begin{hscode}\SaveRestoreHook
\column{B}{@{}>{\hspre}l<{\hspost}@{}}%
\column{5}{@{}>{\hspre}l<{\hspost}@{}}%
\column{E}{@{}>{\hspre}l<{\hspost}@{}}%
\>[B]{}\Varid{benchControl}\mathbin{::}\Conid{Int}\to \Conid{IO}\;(){}\<[E]%
\\
\>[B]{}\Varid{benchControl}\;\Varid{n}\mathrel{=}\mathbf{do}{}\<[E]%
\\
\>[B]{}\hsindent{5}{}\<[5]%
\>[5]{}\Varid{nodes}\leftarrow \Varid{sequence}\mathbin{\circ}\Varid{replicate}\;\Varid{n}\mathbin{\$}\Varid{forkIO}\;(\Varid{return}\;()){}\<[E]%
\\
\>[B]{}\hsindent{5}{}\<[5]%
\>[5]{}\Varid{mapM\char95 }\;\Varid{killThread}\;\Varid{nodes}{}\<[E]%
\ColumnHook
\end{hscode}\resethooks

\subsection{Channel benchmark implementation}
\label{apx:channel-bench-impl}

Each node has references to a send-channel and a receive-channel in the
channel-based implementation.
We reuse the message types from before via an \text{\ttfamily Either}.
\begin{hscode}\SaveRestoreHook
\column{B}{@{}>{\hspre}l<{\hspost}@{}}%
\column{E}{@{}>{\hspre}l<{\hspost}@{}}%
\>[B]{}\mathbf{type}\;\Conid{ChMsg}\mathrel{=}\Conid{Either}\;\Conid{Msg}\;\Conid{Winner}{}\<[E]%
\\
\>[B]{}\mathbf{type}\;\Conid{Ch}\mathrel{=}\Conid{\Conid{Ch}.Chan}\;\Conid{ChMsg}{}\<[E]%
\ColumnHook
\end{hscode}\resethooks

\noindent
It is unnecessary to split the channel-based implementation into a simple node
and an extended node, but we split them anyway to ease comparison to the
actor-based implementation.
This structural similarity hopefully has the added benefit of focusing
benchmark differences onto the communication mechanisms instead of anecdotal
differences.

In \text{\ttfamily chanNode} we implement the main loop.
The only state maintained is the greatest nominee seen.
It leaves off with definitions of communication functions in its where-clause.
\begin{hscode}\SaveRestoreHook
\column{B}{@{}>{\hspre}l<{\hspost}@{}}%
\column{3}{@{}>{\hspre}l<{\hspost}@{}}%
\column{5}{@{}>{\hspre}l<{\hspost}@{}}%
\column{E}{@{}>{\hspre}l<{\hspost}@{}}%
\>[B]{}\Varid{chanNode}\mathbin{::}{}\<[E]%
\\
\>[B]{}\hsindent{5}{}\<[5]%
\>[5]{}\Conid{\Conid{Mv}.MVar}\;\Conid{ThreadId}\to (\Conid{Ch},\Conid{Ch})\to \Conid{ThreadId}\to \Conid{IO}\;(){}\<[E]%
\\
\>[B]{}\Varid{chanNode}\;\Varid{done}\;\Varid{chans}\;\Varid{st}\mathrel{=}\mathbf{do}{}\<[E]%
\\
\>[B]{}\hsindent{5}{}\<[5]%
\>[5]{}\Varid{chanNode}\;\Varid{done}\;\Varid{chans}\mathbin{=\hspace{-0.3em}<\hspace{-0.4em}<}\Varid{exnodePart}\;\Varid{st}\mathbin{=\hspace{-0.3em}<\hspace{-0.4em}<}\Varid{recv}{}\<[E]%
\\
\>[B]{}\hsindent{3}{}\<[3]%
\>[3]{}\mathbf{where}{}\<[E]%
\\
\>[3]{}\hsindent{2}{}\<[5]%
\>[5]{}\Varid{recv}\mathrel{=}\Varid{\Conid{Ch}.readChan}\;(\Varid{fst}\;\Varid{chans}){}\<[E]%
\\
\>[3]{}\hsindent{2}{}\<[5]%
\>[5]{}\Varid{sendMsg}\mathrel{=}\Varid{\Conid{Ch}.writeChan}\;(\Varid{snd}\;\Varid{chans})\mathbin{\circ}\Conid{Left}{}\<[E]%
\\
\>[3]{}\hsindent{2}{}\<[5]%
\>[5]{}\Varid{sendWinner}\mathrel{=}\Varid{\Conid{Ch}.writeChan}\;(\Varid{snd}\;\Varid{chans})\mathbin{\circ}\Conid{Right}{}\<[E]%
\ColumnHook
\end{hscode}\resethooks

\noindent
Within the where-clause of \text{\ttfamily chanNode},
we define \text{\ttfamily nodePart}
to implement the behavior of a ring node from \Cref{sec:ring-intent-fun}.
This part has no state and requires no \text{\ttfamily Init} message.
\begin{hscode}\SaveRestoreHook
\column{B}{@{}>{\hspre}l<{\hspost}@{}}%
\column{5}{@{}>{\hspre}l<{\hspost}@{}}%
\column{9}{@{}>{\hspre}l<{\hspost}@{}}%
\column{10}{@{}>{\hspre}l<{\hspost}@{}}%
\column{13}{@{}>{\hspre}l<{\hspost}@{}}%
\column{17}{@{}>{\hspre}l<{\hspost}@{}}%
\column{23}{@{}>{\hspre}l<{\hspost}@{}}%
\column{31}{@{}>{\hspre}l<{\hspost}@{}}%
\column{E}{@{}>{\hspre}l<{\hspost}@{}}%
\>[5]{}\Varid{nodePart}\mathbin{::}\Conid{Msg}\to \Conid{IO}\;(){}\<[E]%
\\
\>[5]{}\Varid{nodePart}\;\Conid{Start}\mathrel{=}\mathbf{do}{}\<[E]%
\\
\>[5]{}\hsindent{4}{}\<[9]%
\>[9]{}\Varid{self}\leftarrow \Varid{myThreadId}{}\<[E]%
\\
\>[5]{}\hsindent{4}{}\<[9]%
\>[9]{}\Varid{putStrLn}\;(\Varid{show}\;\Varid{self}\mathbin{+\hspace{-0.2em}+}\text{\ttfamily \char34 :~nominate~self\char34}){}\<[E]%
\\
\>[5]{}\hsindent{4}{}\<[9]%
\>[9]{}\Varid{sendMsg}\mathbin{\$}\Conid{Nominate}\;\Varid{self}{}\<[E]%
\\
\>[5]{}\Varid{nodePart}\;\Conid{Nominate}\;\!\{\Varid{nominee}\mathrel{=}\Varid{nom}\mskip1.5mu\}\mathrel{=}\mathbf{do}{}\<[E]%
\\
\>[5]{}\hsindent{4}{}\<[9]%
\>[9]{}\Varid{self}\leftarrow \Varid{myThreadId}{}\<[E]%
\\
\>[5]{}\hsindent{4}{}\<[9]%
\>[9]{}\mathbf{case}\;()\;\mathbf{of}{}\<[E]%
\\
\>[9]{}\hsindent{1}{}\<[10]%
\>[10]{}\anonymous {}\<[13]%
\>[13]{}\mid \Varid{self}\equiv \Varid{nom}\to \Varid{putStrLn}\;(\Varid{show}\;\Varid{self}\mathbin{+\hspace{-0.2em}+}\text{\ttfamily \char34 :~I~win\char34}){}\<[E]%
\\
\>[13]{}\mid \Varid{self}\mathbin{<}{}\<[23]%
\>[23]{}\Varid{nom}\to {}\<[E]%
\\
\>[13]{}\hsindent{4}{}\<[17]%
\>[17]{}\Varid{putStrLn}\;(\Varid{show}\;\Varid{self}\mathbin{+\hspace{-0.2em}+}\text{\ttfamily \char34 :~nominate~\char34}\mathbin{+\hspace{-0.2em}+}\Varid{show}\;\Varid{nom}){}\<[E]%
\\
\>[13]{}\hsindent{4}{}\<[17]%
\>[17]{}\mathbin{>\hspace{-0.4em}>}\Varid{sendMsg}\;(\Conid{Nominate}\;\Varid{nom}){}\<[E]%
\\
\>[13]{}\mid \Varid{otherwise}{}\<[31]%
\>[31]{}\to \Varid{putStrLn}\;\text{\ttfamily \char34 Ignored~nominee\char34}{}\<[E]%
\ColumnHook
\end{hscode}\resethooks
\ignore{
\begin{hscode}\SaveRestoreHook
\column{B}{@{}>{\hspre}l<{\hspost}@{}}%
\column{5}{@{}>{\hspre}l<{\hspost}@{}}%
\column{E}{@{}>{\hspre}l<{\hspost}@{}}%
\>[5]{}\Varid{nodePart}\;\anonymous \mathrel{=}\Varid{error}\;\text{\ttfamily \char34 nodePart:~unhandled\char34}{}\<[E]%
\ColumnHook
\end{hscode}\resethooks
}

\noindent
Still within the where-clause of \texttt{chanNode}, we implement
\texttt{exnodePart} with the behavior of the winner-round node
(\Cref{sec:ring2-intent-fun}) and the benchmark-node
(\Cref{apx:actor-bench-impl}).
It signals termination by placing the confirmed winner's \texttt{ThreadId} into
an \texttt{MVar}.
\begin{hscode}\SaveRestoreHook
\column{B}{@{}>{\hspre}l<{\hspost}@{}}%
\column{5}{@{}>{\hspre}l<{\hspost}@{}}%
\column{9}{@{}>{\hspre}l<{\hspost}@{}}%
\column{13}{@{}>{\hspre}l<{\hspost}@{}}%
\column{17}{@{}>{\hspre}l<{\hspost}@{}}%
\column{21}{@{}>{\hspre}l<{\hspost}@{}}%
\column{E}{@{}>{\hspre}l<{\hspost}@{}}%
\>[5]{}\Varid{exnodePart}\mathbin{::}\Conid{ThreadId}\to \Conid{Either}\;\Conid{Msg}\;\Conid{Winner}\to \Conid{IO}\;\Conid{ThreadId}{}\<[E]%
\\
\>[5]{}\Varid{exnodePart}\;\Varid{great}\;(\Conid{Left}\;\Varid{m})\mathrel{=}\mathbf{do}{}\<[E]%
\\
\>[5]{}\hsindent{4}{}\<[9]%
\>[9]{}\Varid{nodePart}\;\Varid{m}{}\<[E]%
\\
\>[5]{}\hsindent{4}{}\<[9]%
\>[9]{}\Varid{self}\leftarrow \Varid{myThreadId}{}\<[E]%
\\
\>[5]{}\hsindent{4}{}\<[9]%
\>[9]{}\mathbf{case}\;\Varid{m}\;\mathbf{of}{}\<[E]%
\\
\>[9]{}\hsindent{4}{}\<[13]%
\>[13]{}\Conid{Nominate}\;\!\{\Varid{nominee}\mskip1.5mu\}\to {}\<[E]%
\\
\>[13]{}\hsindent{4}{}\<[17]%
\>[17]{}\mathbf{if}\;\Varid{self}\equiv \Varid{nominee}{}\<[E]%
\\
\>[13]{}\hsindent{4}{}\<[17]%
\>[17]{}\mathbf{then}\;\Varid{sendWinner}\;(\Conid{Winner}\;\Varid{self}){}\<[E]%
\\
\>[17]{}\hsindent{4}{}\<[21]%
\>[21]{}\mathbin{>\hspace{-0.4em}>}\Varid{return}\;\Varid{great}{}\<[E]%
\\
\>[13]{}\hsindent{4}{}\<[17]%
\>[17]{}\mathbf{else}\;\Varid{return}\mathbin{\$}\Varid{max}\;\Varid{nominee}\;\Varid{great}{}\<[E]%
\\
\>[9]{}\hsindent{4}{}\<[13]%
\>[13]{}\anonymous \to \Varid{return}\;\Varid{great}{}\<[E]%
\\
\>[5]{}\Varid{exnodePart}\;\Varid{great}\;(\Conid{Right}\;\Varid{m})\mathrel{=}\mathbf{do}{}\<[E]%
\\
\>[5]{}\hsindent{4}{}\<[9]%
\>[9]{}\Varid{self}\leftarrow \Varid{myThreadId}{}\<[E]%
\\
\>[5]{}\hsindent{4}{}\<[9]%
\>[9]{}\mathbf{case}\;\Varid{m}\;\mathbf{of}{}\<[E]%
\\
\>[9]{}\hsindent{4}{}\<[13]%
\>[13]{}\Conid{Winner}\;\Varid{w}{}\<[E]%
\\
\>[13]{}\hsindent{4}{}\<[17]%
\>[17]{}\mid \Varid{w}\equiv \Varid{self}\to {}\<[E]%
\\
\>[17]{}\hsindent{4}{}\<[21]%
\>[21]{}\Varid{putStrLn}\;(\Varid{show}\;\Varid{self}\mathbin{+\hspace{-0.2em}+}\text{\ttfamily \char34 :~Confirmed\char34}){}\<[E]%
\\
\>[17]{}\hsindent{4}{}\<[21]%
\>[21]{}\mathbin{>\hspace{-0.4em}>}\Varid{\Conid{Mv}.putMVar}\;\Varid{done}\;\Varid{self}{}\<[E]%
\\
\>[13]{}\hsindent{4}{}\<[17]%
\>[17]{}\mid \Varid{w}\equiv \Varid{great}\to \Varid{sendWinner}\;(\Conid{Winner}\;\Varid{w}){}\<[E]%
\\
\>[13]{}\hsindent{4}{}\<[17]%
\>[17]{}\mid \Varid{otherwise}\to \Varid{putStrLn}\;\text{\ttfamily \char34 Unexpected~winner\char34}{}\<[E]%
\\
\>[5]{}\hsindent{4}{}\<[9]%
\>[9]{}\Varid{return}\;\Varid{great}{}\<[E]%
\ColumnHook
\end{hscode}\resethooks

\noindent
Finally, we initialize the algorithm with a function similar to
\texttt{ringElection}, but using channels instead of passing in
\texttt{ThreadId}s.
(1) Define a function to run a channel-node on the ``done'' \texttt{MVar} and
two provided channels.
(2) Construct channels and a ring of un-evaluated nodes \emph{in order}.
(3) Finally permute the nodes and fork them out of order.
Nodes are assigned random thread identifiers at this point.
(4) Start the election.
(5) Wait for termination and clean up.
\begin{hscode}\SaveRestoreHook
\column{B}{@{}>{\hspre}l<{\hspost}@{}}%
\column{5}{@{}>{\hspre}l<{\hspost}@{}}%
\column{13}{@{}>{\hspre}l<{\hspost}@{}}%
\column{56}{@{}>{\hspre}l<{\hspost}@{}}%
\column{E}{@{}>{\hspre}l<{\hspost}@{}}%
\>[B]{}\Varid{benchChannels}\mathbin{::}\Conid{Int}\to \Conid{IO}\;(){}\<[E]%
\\
\>[B]{}\Varid{benchChannels}\;\Varid{n}\mathrel{=}\mathbf{do}{}\<[E]%
\\
\>[B]{}\hsindent{5}{}\<[5]%
\>[5]{}\Varid{done}\leftarrow \Varid{\Conid{Mv}.newEmptyMVar}{}\<[E]%
\\
\>[B]{}\hsindent{5}{}\<[5]%
\>[5]{}\mathbf{let}\;\Varid{mkNode}\;\Varid{chans}\mathrel{=}\mathbf{do}\;\hfill(1){}\<[E]%
\\
\>[5]{}\hsindent{8}{}\<[13]%
\>[13]{}\Varid{great}\leftarrow \Varid{myThreadId}{}\<[E]%
\\
\>[5]{}\hsindent{8}{}\<[13]%
\>[13]{}\Varid{chanNode}\;\Varid{done}\;\Varid{chans}\;\Varid{great}{}\<[E]%
\\
\>[B]{}\hsindent{5}{}\<[5]%
\>[5]{}\Varid{chans}\leftarrow \Varid{sequence}\mathbin{\circ}\Varid{replicate}\;\Varid{n}\mathbin{\$}\Varid{\Conid{Ch}.newChan}{}\<[E]%
\\
\>[B]{}\hsindent{5}{}\<[5]%
\>[5]{}\mathbf{let}\;\Varid{nodeActs}\mathrel{=}\Varid{map}\;\Varid{mkNode}\hfill(2){}\<[E]%
\\
\>[5]{}\hsindent{8}{}\<[13]%
\>[13]{}(\Varid{zip}\;\Varid{chans}\mathbin{\$}\Varid{tail}\;\Varid{chans}\mathbin{+\hspace{-0.2em}+}[\Varid{head}\;\Varid{chans}\mskip1.5mu]){}\<[E]%
\\
\>[B]{}\hsindent{5}{}\<[5]%
\>[5]{}\Varid{ringActs}\leftarrow \Varid{getStdRandom}\mathbin{\$}\Varid{permute}\;\Varid{nodeActs}{}\<[E]%
\\
\>[B]{}\hsindent{5}{}\<[5]%
\>[5]{}\Varid{ring}\leftarrow \Varid{mapM}\;\Varid{forkIO}\;\Varid{ringActs}\hfill(3){}\<[E]%
\\
\>[B]{}\hsindent{5}{}\<[5]%
\>[5]{}\Varid{mapM\char95 }\;(\lambda \Varid{c}\to \Varid{\Conid{Ch}.writeChan}\;\Varid{c}\mathbin{\circ}\Conid{Left}\mathbin{\$}\Conid{Start})\;\Varid{chans}{}\<[56]%
\>[56]{}\quad\hfill(4){}\<[E]%
\\
\>[B]{}\hsindent{5}{}\<[5]%
\>[5]{}\Varid{w}\leftarrow \Varid{\Conid{Mv}.takeMVar}\;\Varid{done}\hfill(5){}\<[E]%
\\
\>[B]{}\hsindent{5}{}\<[5]%
\>[5]{}\Varid{mapM\char95 }\;\Varid{killThread}\;\Varid{ring}{}\<[E]%
\\
\>[B]{}\hsindent{5}{}\<[5]%
\>[5]{}\Varid{assert}\;(\Varid{w}\equiv \Varid{maximum}\;\Varid{ring})\;(\Varid{return}\;()){}\<[E]%
\ColumnHook
\end{hscode}\resethooks

\subsection{Criterion benchmark implementation}
\label{apx:criterion-bench-impl}

Finally, we define a benchmark-heat to run each of the benchmark functions
defined above for a given ring size.
The main function (not shown) calls \text{\ttfamily benchHeat} and passes it to
Criterion's \text{\ttfamily defaultMain}.
\begin{hscode}\SaveRestoreHook
\column{B}{@{}>{\hspre}l<{\hspost}@{}}%
\column{5}{@{}>{\hspre}l<{\hspost}@{}}%
\column{E}{@{}>{\hspre}l<{\hspost}@{}}%
\>[B]{}\Varid{benchHeat}\mathbin{::}\Conid{Int}\to \Conid{\Conid{Cr}.Benchmark}{}\<[E]%
\\
\>[B]{}\Varid{benchHeat}\;\Varid{n}\mathrel{=}\Varid{\Conid{Cr}.bgroup}\;(\text{\ttfamily \char34 n=\char34}\mathbin{+\hspace{-0.2em}+}\Varid{show}\;\Varid{n}){}\<[E]%
\\
\>[B]{}\hsindent{5}{}\<[5]%
\>[5]{}[\Varid{\Conid{Cr}.bench}\;\text{\ttfamily \char34 control\char34}\mathbin{\circ}\Varid{\Conid{Cr}.nfIO}\mathbin{\$}\Varid{benchControl}\;\Varid{n}{}\<[E]%
\\
\>[B]{}\hsindent{5}{}\<[5]%
\>[5]{},\Varid{\Conid{Cr}.bench}\;\text{\ttfamily \char34 actor~ring\char34}\mathbin{\circ}\Varid{\Conid{Cr}.nfIO}\mathbin{\$}\Varid{benchActors}\;\Varid{n}{}\<[E]%
\\
\>[B]{}\hsindent{5}{}\<[5]%
\>[5]{},\Varid{\Conid{Cr}.bench}\;\text{\ttfamily \char34 channel~ring\char34}\mathbin{\circ}\Varid{\Conid{Cr}.nfIO}\mathbin{\$}\Varid{benchChannels}\;\Varid{n}\mskip1.5mu]{}\<[E]%
\ColumnHook
\end{hscode}\resethooks

\subsection{Experimental setup and procedure}
\label{apx:exp-setup}

In all benchmarks, we replace printlines with \text{\ttfamily pure~\char40{}\char41{}}
to reduce noise and latency in results.
We compile with the threaded RTS (\text{\ttfamily \char45{}threaded}) and run on all capabilities
(\text{\ttfamily \char43{}RTS~\char45{}N}).
Our test machines included:
\begin{itemize}[leftmargin=1em]
    \item[--] MacBookPro11,5; 8 capabilities (NixOS).
    \item[--] AWS \text{\ttfamily c3\char46{}8xlarge}; 32 vCPU (Amazon Linux 2023 AMI).
    \item[--] AWS \text{\ttfamily c6a\char46{}48xlarge}; 192 vCPU (Amazon Linux 2023 AMI).
\end{itemize}
Our experiment proceeded as follows:
\begin{itemize}[leftmargin=1em]
    \item[--] We ran the \text{\ttfamily criterion} benchmark for ring sizes up to
    $16384$ on the MacBookPro11,5, clocked to 1.6GHz, without
    frequency scaling, and with no other programs running (kernel vtty).
    Channels took a third the time of actors, but
    the performance gap narrowed as ring size increased.

    \item[--] \Cref{fig:perf-eval-time-n32}:
    We ran the benchmark on the AWS \text{\ttfamily c3\char46{}8xlarge} instance with 32 vCPU
    for ring sizes up to $65536$.
    We saw actors outperform channels at high ring sizes.

    \item[--] \Cref{fig:perf-eval-time-n192}:
    We ran the benchmark on the AWS \text{\ttfamily c6a\char46{}48xlarge} instance with 192 vCPU
    for ring sizes up to $65536$.
    The benchmark segfaulted unpredictably.
    We used a shell script to call the benchmark executable once per set of
    parameters to work around segfaults.
    We confirmed that actors outperform channels at high ring sizes.

    \item[--] \Cref{fig:perf-eval-time-n8}:
    We repeated the benchmark on the MacBookPro11,5
    for ring sizes up to $65536$.

    \item[--] \Cref{fig:perf-eval-mem}:
    We ran a different benchmark
    to measure total-allocations (\text{\ttfamily \char43{}RTS~\char45{}t~\char45{}\char45{}machine\char45{}readable})
    on the MacBookPro11,5
    for ring sizes up to $65536$.
    Here the main function only ran a single algorithm at a
    specified ring size once, and then terminated.
    We ran ten trials for each combination of algorithm and ring size, and averaged.
\end{itemize}

\subsection{Experiment result}
\label{apx:exp-result}

\begin{figure}
    \begin{subfigure}{\linewidth}
    {\small
        \def\svgwidth{\linewidth}
        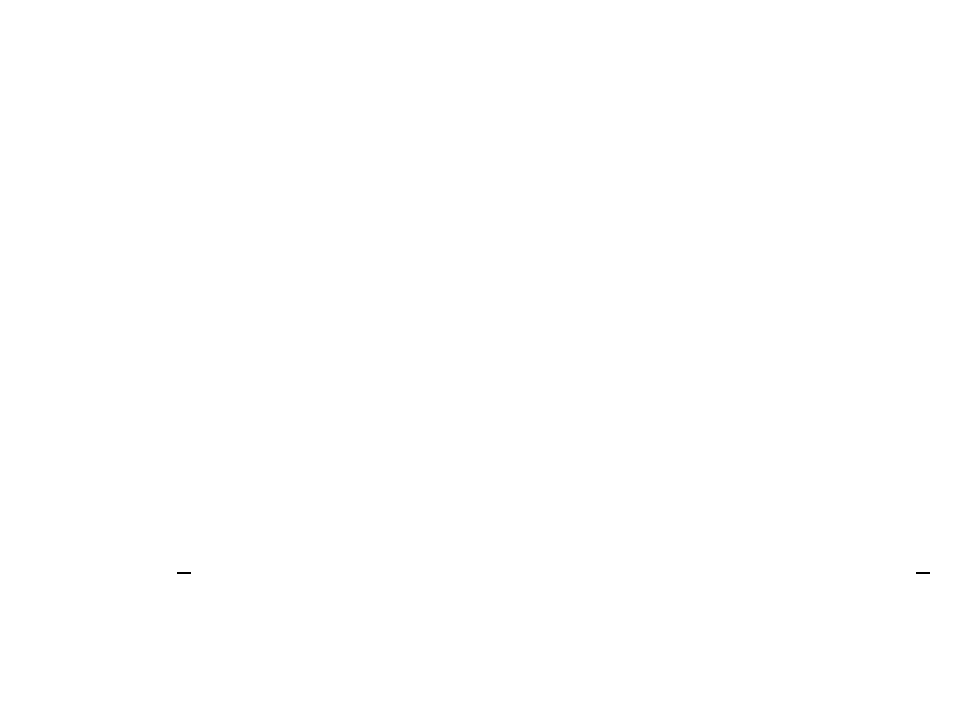
    }
    \caption{Running time with 8 capabilities.}
    \label{fig:perf-eval-time-n8}
    \end{subfigure}

    \begin{subfigure}{\linewidth}
    {\small
        \def\svgwidth{\linewidth}
        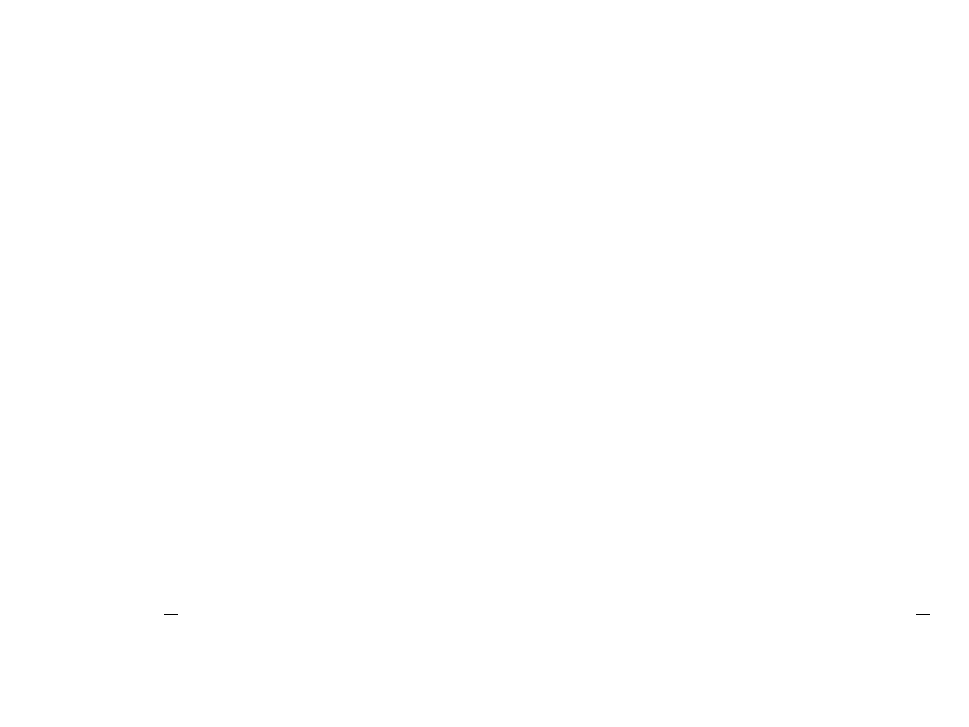
    }
    \caption{
        Running time with 192 capabilities:
        The missing datapoint is a run that consistently crashed with a segmentation fault in the RTS.
    }
    \label{fig:perf-eval-time-n192}
    \end{subfigure}

\caption{
    On different machines we replicate
    \Cref{fig:perf-eval-time-n32} in both the absolute running time, and
    the actor-based implementation being faster at the highest
    ring sizes.
}
\label{fig:perf-eval-time-rest}
\end{figure}

\begin{figure}
    {\small
        \def\svgwidth{\linewidth}
        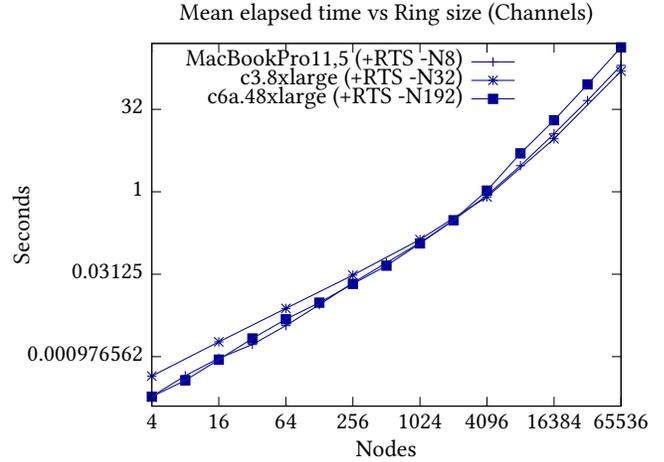
    }
    \caption{
        The growth of total-allocations as ring size is increased inflects to a
        higher rate near $2048$ nodes.
    }
    \label{fig:perf-group-chan}
\end{figure}

Our running time results for 8, 32, and 192 capabilities are in
\Cref{fig:perf-eval-time-n8,fig:perf-eval-time-n32,fig:perf-eval-time-n192},
respectively.
We group the running time of the channel-based implementation over all three
machines in \Cref{fig:perf-group-chan} to make its inflection point clearer.
Our total-allocations result is in \Cref{fig:perf-eval-mem}.

The running time of the extended ring leader election is $O(2n)$ in the number
of nodes.
We hypothesize that it is invariant to the number of capabilities because after
an initial flood of nominations, the algorithm degenerates quickly to a single
message passing around the ring twice.

\subsection{Actor-based (dynamic types) trace}
\label{apx:main2-trace}

In \Cref{sec:main2-init}, we showed how to call \text{\ttfamily runElection} on
\text{\ttfamily exnode} to run a ring leader election with a winner declaration round.
Here is an example trace.

\scriptsize

\begin{tabbing}\ttfamily
~\char62{}~main2~4\\
\ttfamily ~ThreadId~53~send~Init~\char123{}next~\char61{}~ThreadId~57\char125{}~to~ThreadId~55\\
\ttfamily ~ThreadId~53~send~Init~\char123{}next~\char61{}~ThreadId~54\char125{}~to~ThreadId~57\\
\ttfamily ~ThreadId~53~send~Init~\char123{}next~\char61{}~ThreadId~56\char125{}~to~ThreadId~54\\
\ttfamily ~ThreadId~53~send~Init~\char123{}next~\char61{}~ThreadId~55\char125{}~to~ThreadId~56\\
\ttfamily ~ThreadId~53~send~Start~to~ThreadId~55\\
\ttfamily ~ThreadId~55~send~Nominate~\char123{}nominee~\char61{}~ThreadId~55\char125{}~to~ThreadId~57\\
\ttfamily ~ThreadId~53~send~Start~to~ThreadId~57\\
\ttfamily ~Ignored~nomination\\
\ttfamily ~ThreadId~57~send~Nominate~\char123{}nominee~\char61{}~ThreadId~57\char125{}~to~ThreadId~54\\
\ttfamily ~ThreadId~53~send~Start~to~ThreadId~54\\
\ttfamily ~ThreadId~54~send~Nominate~\char123{}nominee~\char61{}~ThreadId~57\char125{}~to~ThreadId~56\\
\ttfamily ~ThreadId~56~send~Nominate~\char123{}nominee~\char61{}~ThreadId~57\char125{}~to~ThreadId~55\\
\ttfamily ~ThreadId~54~send~Nominate~\char123{}nominee~\char61{}~ThreadId~54\char125{}~to~ThreadId~56\\
\ttfamily ~ThreadId~53~send~Start~to~ThreadId~56\\
\ttfamily ~ThreadId~55~send~Nominate~\char123{}nominee~\char61{}~ThreadId~57\char125{}~to~ThreadId~57\\
\ttfamily ~Ignored~nomination\\
\ttfamily ~ThreadId~57\char58{}~I~win\\
\ttfamily ~ThreadId~57~send~Winner~\char40{}ThreadId~57\char41{}~to~ThreadId~54\\
\ttfamily ~ThreadId~56~send~Nominate~\char123{}nominee~\char61{}~ThreadId~56\char125{}~to~ThreadId~55\\
\ttfamily ~ThreadId~54~send~Winner~\char40{}ThreadId~57\char41{}~to~ThreadId~56\\
\ttfamily ~ThreadId~55~send~Nominate~\char123{}nominee~\char61{}~ThreadId~56\char125{}~to~ThreadId~57\\
\ttfamily ~ThreadId~56~send~Winner~\char40{}ThreadId~57\char41{}~to~ThreadId~55\\
\ttfamily ~Ignored~nomination\\
\ttfamily ~ThreadId~55~send~Winner~\char40{}ThreadId~57\char41{}~to~ThreadId~57\\
\ttfamily ~ThreadId~57\char58{}~Confirmed
\end{tabbing}

\normalsize

\subsection{Channel-based extended election trace}
\label{apx:benchChannels-trace}

In \Cref{apx:control-bench-impl} we defined \text{\ttfamily benchChannels} to run a ring
leader election with a winner declaration round using channels for
communication.
Here's an example trace.

\scriptsize

\begin{tabbing}\ttfamily
~\char62{}~benchChannels~4\\
\ttfamily ~ThreadId~61\char58{}~nominate~self\\
\ttfamily ~ThreadId~62\char58{}~nominate~self\\
\ttfamily ~ThreadId~63\char58{}~nominate~self\\
\ttfamily ~ThreadId~64\char58{}~nominate~self\\
\ttfamily ~Ignored~nominee\\
\ttfamily ~ThreadId~63\char58{}~nominate~ThreadId~64\\
\ttfamily ~ThreadId~61\char58{}~nominate~ThreadId~63\\
\ttfamily ~ThreadId~61\char58{}~nominate~ThreadId~64\\
\ttfamily ~Ignored~nominee\\
\ttfamily ~ThreadId~62\char58{}~nominate~ThreadId~63\\
\ttfamily ~ThreadId~62\char58{}~nominate~ThreadId~64\\
\ttfamily ~Ignored~nominee\\
\ttfamily ~ThreadId~64\char58{}~I~win\\
\ttfamily ~ThreadId~64\char58{}~Confirmed
\end{tabbing}

\normalsize

\ignore{
\begin{hscode}\SaveRestoreHook
\column{B}{@{}>{\hspre}l<{\hspost}@{}}%
\column{3}{@{}>{\hspre}l<{\hspost}@{}}%
\column{5}{@{}>{\hspre}l<{\hspost}@{}}%
\column{9}{@{}>{\hspre}l<{\hspost}@{}}%
\column{13}{@{}>{\hspre}l<{\hspost}@{}}%
\column{E}{@{}>{\hspre}l<{\hspost}@{}}%
\>[B]{}\Varid{beginVerb}\mathbin{::}\Conid{IO}\;(){}\<[E]%
\\
\>[B]{}\Varid{beginVerb}\mathrel{=}\mathbf{do}{}\<[E]%
\\
\>[B]{}\hsindent{5}{}\<[5]%
\>[5]{}\Varid{hSetBuffering}\;\Varid{stdout}\;\Conid{LineBuffering}{}\<[E]%
\\
\>[B]{}\hsindent{5}{}\<[5]%
\>[5]{}\Varid{putStrLn}\;\text{\ttfamily \char34 \char92 \char92 begin\char123 verbatim\char125 \char34}{}\<[E]%
\\[\blanklineskip]%
\>[B]{}\Varid{endVerb}\mathbin{::}\Conid{IO}\;(){}\<[E]%
\\
\>[B]{}\Varid{endVerb}\mathrel{=}\Varid{putStrLn}\;\text{\ttfamily \char34 \char92 \char92 end\char123 verbatim\char125 \char34}{}\<[E]%
\\[\blanklineskip]%
\>[B]{}\Varid{main}\mathbin{::}\Conid{IO}\;(){}\<[E]%
\\
\>[B]{}\Varid{main}\mathrel{=}\mathbf{do}{}\<[E]%
\\
\>[B]{}\hsindent{5}{}\<[5]%
\>[5]{}\Varid{ringSize}\leftarrow \Varid{maybe}\;\mathrm{8}\;\Varid{read}\mathbin{`\Varid{fmap}`}\Varid{lookupEnv}\;\text{\ttfamily \char34 RING\char95 SIZE\char34}{}\<[E]%
\\
\>[B]{}\hsindent{5}{}\<[5]%
\>[5]{}\Varid{modeRaw}\leftarrow \Varid{lookupEnv}\;\text{\ttfamily \char34 MODE\char34}{}\<[E]%
\\
\>[B]{}\hsindent{5}{}\<[5]%
\>[5]{}\Varid{print}\;(\text{\ttfamily \char34 RING\char95 SIZE\char34},\Varid{ringSize},\text{\ttfamily \char34 a~natural~number\char34}){}\<[E]%
\\
\>[B]{}\hsindent{5}{}\<[5]%
\>[5]{}\Varid{print}\;(\text{\ttfamily \char34 MODE\char34},\Varid{modeRaw},\text{\ttfamily \char34 actors~|~channels~|~control~|~<UNSET:criterion>\char34}){}\<[E]%
\\
\>[B]{}\hsindent{5}{}\<[5]%
\>[5]{}\mathbf{case}\;\Varid{modeRaw}\;\mathbf{of}{}\<[E]%
\\
\>[5]{}\hsindent{4}{}\<[9]%
\>[9]{}\Conid{Just}\;\text{\ttfamily \char34 actors\char34}\to \mathbf{do}{}\<[E]%
\\
\>[9]{}\hsindent{4}{}\<[13]%
\>[13]{}\Varid{putStrLn}\;\text{\ttfamily \char34 benchActors~function\char34}{}\<[E]%
\\
\>[9]{}\hsindent{4}{}\<[13]%
\>[13]{}\Varid{benchActors}\;\Varid{ringSize}{}\<[E]%
\\
\>[5]{}\hsindent{4}{}\<[9]%
\>[9]{}\Conid{Just}\;\text{\ttfamily \char34 channels\char34}\to \mathbf{do}{}\<[E]%
\\
\>[9]{}\hsindent{4}{}\<[13]%
\>[13]{}\Varid{putStrLn}\;\text{\ttfamily \char34 benchChannels~function\char34}{}\<[E]%
\\
\>[9]{}\hsindent{4}{}\<[13]%
\>[13]{}\Varid{benchChannels}\;\Varid{ringSize}{}\<[E]%
\\
\>[5]{}\hsindent{4}{}\<[9]%
\>[9]{}\Conid{Just}\;\text{\ttfamily \char34 control\char34}\to \mathbf{do}{}\<[E]%
\\
\>[9]{}\hsindent{4}{}\<[13]%
\>[13]{}\Varid{putStrLn}\;\text{\ttfamily \char34 benchControl~function\char34}{}\<[E]%
\\
\>[9]{}\hsindent{4}{}\<[13]%
\>[13]{}\Varid{benchControl}\;\Varid{ringSize}{}\<[E]%
\\
\>[5]{}\hsindent{4}{}\<[9]%
\>[9]{}\Conid{Nothing}\to \mathbf{do}{}\<[E]%
\\
\>[9]{}\hsindent{4}{}\<[13]%
\>[13]{}\Varid{putStrLn}\;\text{\ttfamily \char34 criterion~defaultMain\char34}{}\<[E]%
\\
\>[9]{}\hsindent{4}{}\<[13]%
\>[13]{}\Varid{\Conid{Cr}.defaultMain}\;[\Varid{benchHeat}\;\Varid{ringSize}\mskip1.5mu]{}\<[E]%
\\
\>[5]{}\hsindent{4}{}\<[9]%
\>[9]{}\Conid{Just}\;\anonymous \to \Varid{error}\;\text{\ttfamily \char34 unexpected~mode\char34}{}\<[E]%
\\
\>[B]{}\hsindent{3}{}\<[3]%
\>[3]{}\mathbf{where}{}\<[E]%
\\
\>[3]{}\hsindent{2}{}\<[5]%
\>[5]{}\mbox{\onelinecomment  Used for trace figures}{}\<[E]%
\\
\>[3]{}\hsindent{2}{}\<[5]%
\>[5]{}\anonymous \mathrel{=}\Varid{beginVerb}{}\<[E]%
\\
\>[3]{}\hsindent{2}{}\<[5]%
\>[5]{}\anonymous \mathrel{=}\Varid{endVerb}{}\<[E]%
\\
\>[3]{}\hsindent{2}{}\<[5]%
\>[5]{}\anonymous \mathrel{=}\Varid{main1}{}\<[E]%
\\
\>[3]{}\hsindent{2}{}\<[5]%
\>[5]{}\anonymous \mathrel{=}\Varid{main2}{}\<[E]%
\ColumnHook
\end{hscode}\resethooks
}

\end{document}